\documentclass[twocolumn,floatfix]{aastex631}

\newcommand{\ra}[1]{\renewcommand{\arraystretch}{#1}}
\newcommand{\Msol}{\hbox{M$_\sun$}}
\newcommand{\Pa}{Paper~1}

\usepackage{booktabs}

\usepackage{float}
\usepackage{gensymb}
\usepackage{array,multirow,graphicx}

\shorttitle{Lensed Binary System}
\shortauthors{Williams et al.}

\graphicspath{{./}{}}

\begin{document}

\title{JWST's PEARLS: A Candidate Massive Binary Star System in a Lensed Galaxy at Redshift 0.94}

\date{3 July 2025}

\author[0000-0002-1681-0767]{Hayley Williams}
\affiliation{School of Physics and Astronomy, University of Minnesota, 
116 Church Street SE, Minneapolis, MN 55455, USA 
}
\affiliation{School of Earth and Space Exploration, Arizona State University,
Tempe, AZ 85287-6004, USA}

\author[0000-0003-3142-997X]{Patrick L. Kelly}
\affiliation{School of Physics and Astronomy, University of Minnesota, 
116 Church Street SE, Minneapolis, MN 55455, USA 
}

\author[0000-0002-7464-498X]{Emmanouil Zapartas}
\affiliation{Institute of Astrophysics, FORTH, 71110 Heraklion, Greece}

\author[0000-0001-8156-6281]{Rogier A. Windhorst}
\affiliation{School of Earth and Space Exploration, Arizona State University,
Tempe, AZ 85287-6004, USA}
% \email{Rogier.Windhorst@asu.edu}

\author[0000-0003-1949-7638]{Christopher J.\ Conselice} %%% conselice@manchester.ac.uk
\affiliation{Jodrell Bank Centre for Astrophysics, Alan Turing Building,
University of Manchester, Oxford Road, Manchester M13 9PL, UK}

\author[0000-0003-3329-1337]{Seth H. Cohen} %%% seth.cohen@asu.edu
\affiliation{School of Earth and Space Exploration, Arizona State University,
Tempe, AZ 85287-1404, USA}

\author[0000-0003-0203-3853]{Birendra Dhanasingham}
\affiliation{School of Physics and Astronomy, University of Minnesota, 
116 Church Street SE, Minneapolis, MN 55455, USA 
}

\author[0000-0001-9065-3926]{Jos\'e M. Diego}
\affiliation{IFCA, Instituto de F\'isica de Cantabria (UC-CSIC), Av.\ de Los Castros s/n, 39005 Santander, Spain}

\author[0000-0003-3460-0103]{Alexei V. Filippenko}
\affiliation{Department of Astronomy, University of California, Berkeley, CA 94720-3411, USA}

\author[0000-0003-1625-8009]{Brenda L. Frye}
\affiliation{Department of Astronomy/Steward Observatory, University of Arizona, 933 N. Cherry Avenue, Tucson, AZ 85721, USA}

\author[0000-0002-4884-6756]{Benne W. Holwerda}
\affiliation{Department of Physics, University of Louisville, Natural Science Building 102, Louisville, KY 40292, USA}
%\email{benne.holwerda@louisville.edu}

\author[0000-0002-8716-6980]{Terry J. Jones}
\affiliation{School of Physics and Astronomy, University of Minnesota, 
116 Church Street SE, Minneapolis, MN 55455, USA 
}

\author[0000-0002-6610-2048]{Anton M. Koekemoer}
\affiliation{Space Telescope Science Institute, 3700 San Martin Drive,
Baltimore, MD 21218, USA}

\author[0000-0002-7876-4321]{Ashish Kumar Meena}
\affiliation{Physics Department, Ben-Gurion University of the Negev, P.O. Box 653, Be’er-Sheva 84105, Israel}

\author[0000-0003-4223-7324]{Massimo Ricotti}
\affiliation{Department of Astronomy, University of Maryland, College Park, 20742, USA}

\author[0000-0002-5404-1372]{Clayton D. Robertson}
\affiliation{Department of Physics, University of Louisville, Natural Science Building 102, Louisville, KY 40292, USA}
%\email{clayton.robertson.1@louisville.edu}

\author[0000-0002-5319-6620]{Payaswini Saikia}
\affiliation{Center for Astrophysics and Space Science (CASS), New York University Abu Dhabi, P.O. Box 129188, Abu Dhabi, UAE}

\author[0000-0001-7957-6202]{Bangzheng Sun}
\affiliation{Department of Physics and Astronomy, University of Missouri - Columbia, Columbia, MO 65201, USA}

\author[0000-0002-9895-5758]{S.\ P.\ Willner}
\affiliation{Center for Astrophysics \textbar\ Harvard \& Smithsonian, 60 Garden Street, Cambridge, MA, 02138, USA}
%\email{swillner@cfa.harvard.edu}

\author[0000-0001-7592-7714]{Haojing Yan}
\affiliation{Department of Physics and Astronomy, University of Missouri - Columbia, Columbia, MO 65201, USA}

\author[0000-0002-0350-4488]{Adi Zitrin}
\affiliation{Physics Department, Ben-Gurion University of the Negev, P.O. Box 653, Be’er-Sheva 84105, Israel}

\begin{abstract} Massive stars at cosmological distances can be individually detected during transient microlensing events, when gravitational lensing magnifications may exceed $\mu\approx 1000$. Nine such sources were identified in JWST NIRCam imaging of a single galaxy at redshift $z=0.94$ known as the ``Warhol arc,'' which is mirror-imaged by the galaxy cluster MACS~J0416.1$-$2403. Here we present the discovery of two coincident and well-characterized microlensing events at the same location followed by a third event observed in a single filter approximately 18 months later. The events can be explained by microlensing of a binary star system consisting of a red supergiant ($T\approx 4000$~K) and a B-type ($T\gtrsim13,000$~K) companion star. The timescale of the coincident microlensing events constrains the estimated projected source-plane size to tens of AU. The most likely binary configurations consistent with the observational constraints on the temperature and luminosity of each star are stars with initial masses $M1_{\rm init}=23.6^{+5.3}_{-4.3}$~\Msol\ and an initial mass ratio between the two stars close to unity. A kinematic model that reproduces the observed light curve in all filters gives a relatively small transverse velocity of $\sim 50$~km~s$^{-1}$. This requires the dominant velocity component of several hundreds of km s$^{-1}$ to be roughly parallel to the microcaustic. An alternative possibility would be that the three microlensing events correspond to unrelated stars crossing distinct microcaustics, but this would imply a highly elevated rate of events at their common position, even though no underlying knot is present at the location.
\end{abstract}

\section{Introduction}
\begin{figure*}[ht]
    \centering
    \includegraphics[width=\linewidth]{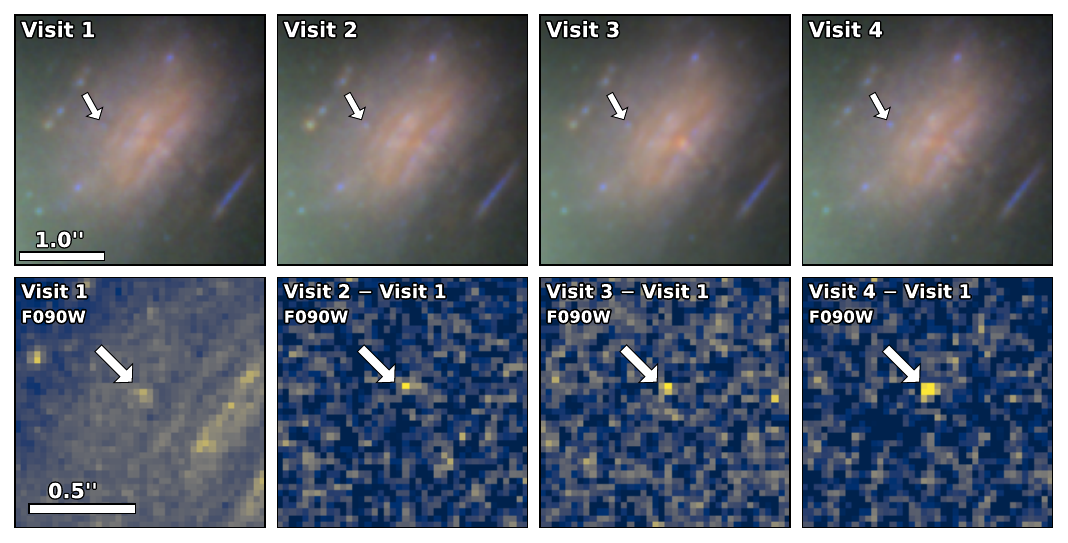}
    \caption{{\it Top:} composite-color images of the Warhol arc the four epochs of JWST NIRCam imaging in which source W2 is detected. {\it Bottom:} Magnified difference images between the epochs in the F090W filter. (The Visit~1 image is the single-epoch image, not a difference image.)  W2 is indicated with a white arrow. North is up and east to the left in all images, and scale bars are provided.}
    \label{fig:cutouts}
\end{figure*}

Binary stellar systems are ubiquitous in nearby galaxies, and binary interactions play a significant role in the evolution of massive stars and their host galaxies \citep[e.g.,][]{Sana_2012}. Binaries are especially common for massive O- and B-type stars, for which the observed multiplicity fraction is at least 50\% and as high as 90\% \citep[e.g.,][]{Sana_2013,Moe_2017,Banyard+2022,Frost_2025,Guo_2022}. OB stars have short main-sequence lifetimes, and evolve into red supergiants (RSGs) with cool temperatures ($3500\lesssim T_{\rm eff}\lesssim4500$ K) and large radii (up to $1500$ R$_\odot$) \citep[e.g.,][]{Humphreys_1979,Levesque_2010,Davies_2013}. The multiplicity fraction of RSGs is lower than that of their progenitors, but binaries still represent a significant portion of the RSG population, 15--40\% \citep[e.g.,][]{Neugent_2020,Patrick_2022,Dai_2025}. 

While thousands of RSG binary systems have been observed and studied in the Milky Way and our closest neighbor galaxies, observational constraints prevent us from detecting and resolving individual star systems at distances greater than $\sim 40$ Mpc. The redshift evolution of the multiplicity fraction and the statistical distribution of binary orbital parameters is unknown, and, given the lack of constraints, the impact of binaries is sometimes not incorporated into stellar population synthesis simulations of high-redshift galaxies. Direct constraints on massive binaries at cosmic noon and beyond would provide important constraints for models of cosmic reionization \citep[e.g.,][]{Gotberg_2020,Doughty_2021},  light-curve diversity of core-collapse SNe \citep[e.g.,][]{Eldridge_2018}, and compact-object mergers \citep[e.g.,][]{Niejsel_2019,Bavera_2021,Klencki_2022,deSa_2024}.  

{Galaxy-cluster gravitational lensing enables the detection of \textit{individual} massive stars at cosmological distances.}
Transient events occur when the magnification of a luminous star lying adjacent to the cluster's critical curve is temporarily greatly boosted by microlensing from an intracluster star or other compact foreground object. The first examples of these sources were discovered with the {Hubble Space Telescope} \citep[HST;][]{Kelly_2018,Rodney_2018,Chen_2019,Kaurov_2019}, and dozens of microlensing events have now been discovered with the {\it James Webb Space Telescope}  \citep[JWST; e.g.,][]{Chen_2019,Kelly_2022,Diego_2023_gordo,Meena_2023,Yan_2023,Fudamoto_2024}.

The redshift $z=0.396$ galaxy cluster MACS J0416.1$-$2304 (hereafter M0416) has been a prolific source of lensed transients \citep[][and references therein]{Yan_2023} including the $z=2.091$ star Mothra \citep{Diego_2023}. Two JWST programs (PEARLS, \citealt{Windhorst_2023}; CANUCS, \citealt{Willott_2022,Sarrouh_2025}), obtained four epochs of NIRCam imaging of M0426 spanning 126 days. {These imaging epochs are hereafter denoted Visits 1, 2, 3, and 4.}  A single lensed galaxy at $z=0.94$, known as the ``Warhol arc,'' produced nine transient events \citep[][]{Yan_2023}. (Williams et al.\ 2025, submitted, hereafter referred to as ``\Pa'') modeled the spectral energy distributions (SEDs) of the Warhol transients with a comprehensive suite of stellar-atmosphere models and found that all nine transients are likely highly magnified RSGs. 
Seven transients were well fit by single stars, one transient showed marginal evidence of being a binary, and one transient (``W2") showed strong evidence for multiplicity, either a binary or small stellar cluster.
The best-fit binary comprised a red component with $\log(T_{\rm eff})=3.55^{+0.06}_{-0.08}$) and a blue component with $\log(T_{\rm eff})=4.10^{+0.23}_{-0.13}$. 
%This paper reports the discovery of a candidate massive binary star system or very small stellar cluster in the Warhol arc at $z=0.94$.  (see Figure \ref{fig:cutouts}). This system was first identified by \cite{Yan_2023}, and is denoted as source W2 in \cite{Yan_2023} and \Pa. The source 

W2 was detected in all four epochs of NIRCam imaging (Figure~\ref{fig:cutouts}), and its flux variation over 126 observer-frame days requires microlensing by a star or other compact object within M0416. This paper analyzes the light curves at the eight NIRCam wavelengths to constrain the nature of the W2 stellar system.
The paper is organized as follows. Section \ref{sec:obs}  describes the JWST/NIRCam observations and the photometry. The source-plane size constraints are presented in Section \ref{sec:size}. Section~\ref{sec:posydon} examines which simulated binary systems are within the observational constraints and analyzes the evolutionary histories of the most likely systems. In Section \ref{sec:orbit}, we model how the magnifications of each star in a binary configuration vary with time as the stars orbit one another across the microlensing caustic. Section \ref{sec:conclusion}  discusses the results and summarizes our conclusions. 
Where distances are needed, they are based on a flat $\Lambda$CDM cosmology with matter-density parameter $\Omega_{M}=0.287$ and Hubble constant H$_0=69.3$~km~s$^{-1}$~Mpc$^{-1}$.

\section{Observations and Photometry}\label{sec:obs}
%PEARLS and CANCUS brief data description
\subsection{Observations}
NIRCam observed the M0416 cluster field in four visits spanning 126 days. Three visits were obtained by the Prime Extragalactic Areas for Reionization and Lensing Science program \citep[PEARLS; ][]{Windhorst_2023}, and one visit was obtained by the CAnadian NIRISS Unbiased Cluster Survey \citep[CANUCS; ][]{Willott_2022}. Both programs imaged the field using the same eight NIRCam filters: F090W, F115W, F150W, F200W, F277W, F356W, F410M, and F444W\null. 
%The exposure times and position angles for each NIRCam visit are listed in Table \ref{tab:obs}.
Table~1 of \Pa\ lists the dates, exposure times, and position angles.

To create images in each filter, we retrieved the Stage 1 PEARLS and CANUCS data products from the Mikulski Archive for Space Telescopes (MAST)\null. {These observations can be accessed via \dataset[doi:10.17909/7rqz-qy32].} The data were reduced using version 1.15.0 of the public JWST science calibration pipeline \citep{bushouse_2023} with reference files from {\tt jwst\_1253.pmap} and all default parameters. Using the Stage~3 JWST pipeline, we resampled the reduced science-level mosaics to a 0\farcs03 pixel scale and projected all epochs onto a common pixel grid.

In addition to the four visits described above, we incorporate NIRCam observations of the M0416 cluster field obtained through the ``Medium-band Astrophysics with the Grism of NIRCam in Frontier Fields'' \citep[MAGNIF; GO-2883, PI F. Sun;][]{2025arXiv250303829F} program in JWST Cycle 2. These observations were carried out using the F210M and F182M filters approximately 6 and 18 months (respectively) after the final PEARLS visit. We also include data from the Cycle 2 program GO-3538 (``Unveiling the Properties of High-Redshift Low/Intermediate-Mass Galaxies in Lensing Fields with NIRCam Wide Field Slitless Spectroscopy''; PI E. Iani), which utilized the F182M filter and was conducted 11 months after the last PEARLS visit. Table~\ref{tab_182_210_flux} summarizes the observation dates, exposure times, and position angles for these additional visits.

\begin{deluxetable*}{ccccccc}
\setlength{\tabcolsep}{6.0pt}
\tablecaption{Additional JWST/NIRCam Observations of M0416} \label{tab_182_210_flux}
\tablehead{
\multirow{2}{*}{MJD} & Proposal & \multirow{2}{*}{Filter} & $t_{\rm exp}$ & P.A. & 5$\sigma$ Detection Limit & \multirow{2}{*}{$F_\nu \,[{\rm \mu} \rm Jy$, AB mag]}\\
& ID & & [s] & [deg] & [${\rm \mu}$Jy, AB mag] \\ [-0.4cm]
}
\startdata
60176 & 2883 & F210M & 9277 & 256.1 &0.0069, 29.3 & $0.0110\pm0.0072$, $28.79^{+1.15}_{-0.55}$\\ 
60326 & 3538 & F182M & 4638 & 53.3 & 0.0088, 29.0&   $0.0130\pm0.0065$, $28.62^{+0.75}_{-0.44}$\\  
60537 & 2883 & F182M & 9277 & 255.5 &  0.0063, 29.4& $0.0150\pm0.0048$, $28.46^{+0.42}_{-0.30}$\\
\enddata
\tablecomments{Columns: Modified Julian Date, JWST proposal ID, filter, exposure time (in seconds), position angle (in degrees), 5 $\sigma$ detection limits (in $\mu$Jy and AB mag), and flux density in $\mu \rm Jy$ and AB mag.}
\end{deluxetable*}

\subsection{Photometry and Light Curve}\label{subsec:photometry}

\begin{figure}[ht]
    \centering
    \includegraphics[width=\linewidth]{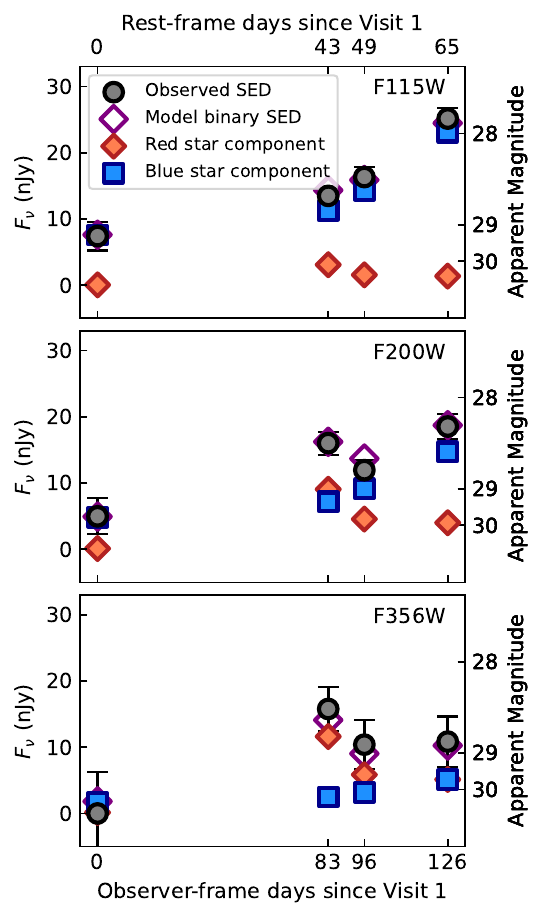}
    \caption{Observed NIRCam light curve of the transient source in the F115W, F200W, and F356W filters (black points). The red diamonds and blue squares show the {\tt POSYDON} evolutionary model flux densities for one of the most likely binary configurations. The open purple diamonds show the sum of the two model binary components.}
    \label{fig:lightcurve}
\end{figure}

\begin{figure*}[ht]
    \includegraphics[width=\linewidth]{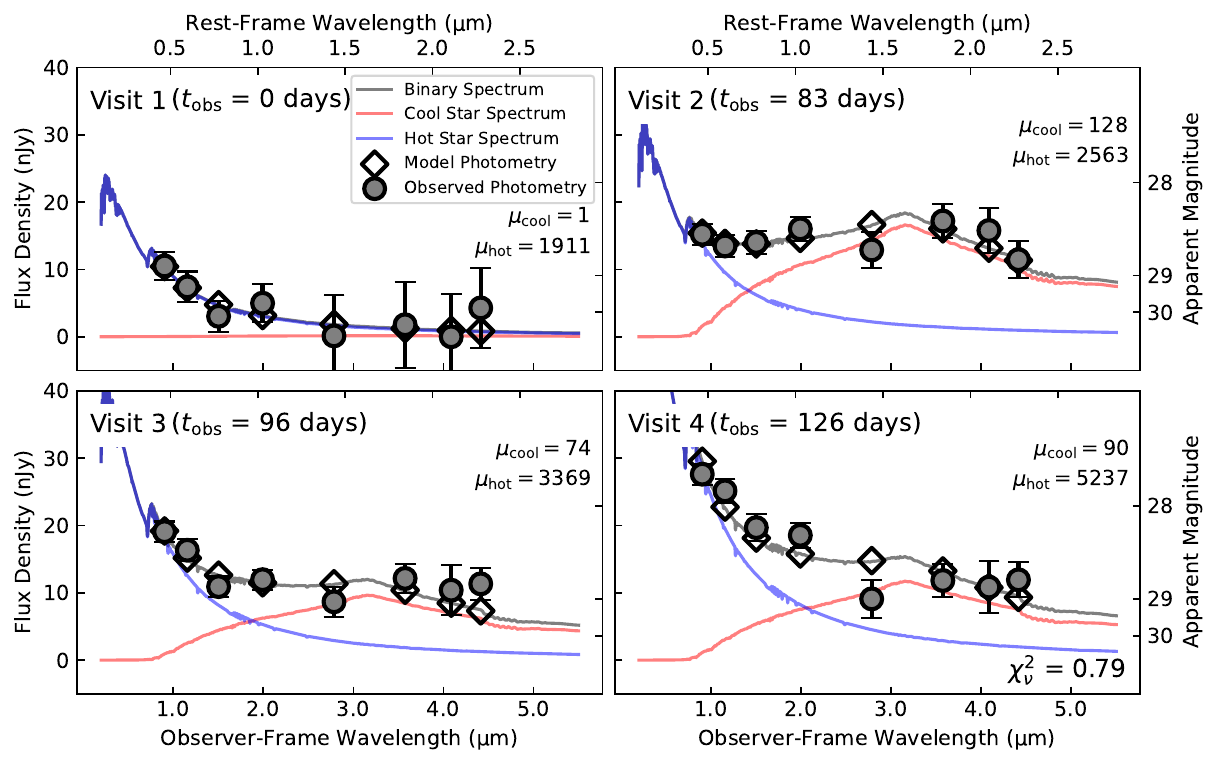}
    \caption{Observed SED of the lensed transient source (black circles). The red and blue lines show the simulated spectra, normalized by their best-fitting magnification factors, for a likely binary configuration from the weighted  {\tt POSYDON} sample. The open black diamonds indicate the synthetic NIRCam photometry of the sum of the two model spectra. }
    \label{fig:SED}
\end{figure*}

\begin{figure}
    \centering
    \includegraphics[width=\linewidth]{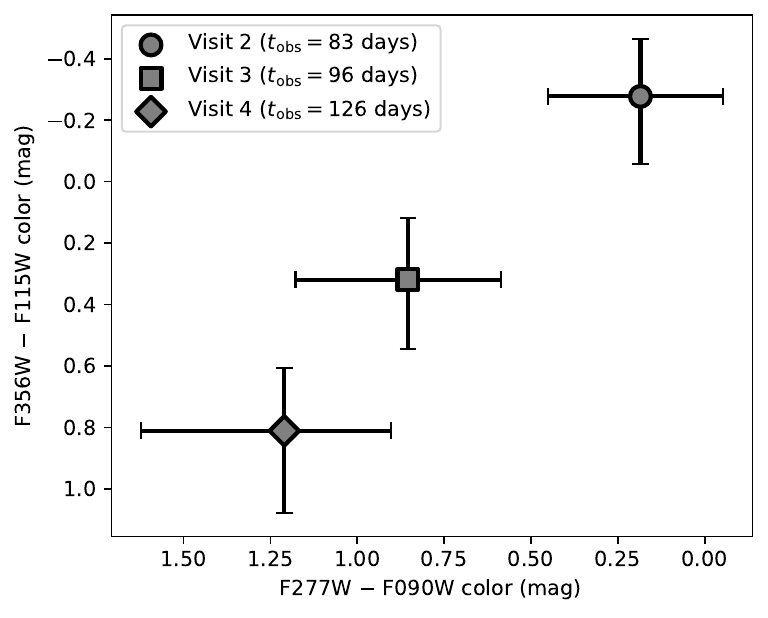}
    \caption{Change in color of the lensed transient over the 126-day light curve for two pairs of filters: F277W$-$F090W and F356W$-$F115W. }
    \label{fig:color_curve}
\end{figure}

Because the transient sources are embedded in the Warhol arc, difference images are necessary to obtain precise differential photometry of each source. As explained in \Pa, W2 was detected in all four epochs, so we used Visit 1 (in which the source was the faintest) as the ``template'' and  created difference images by subtracting the Visit~1  images from the later ones. 
%The photometry is then performed on the Visit 1 images and on the difference images for the subsequent epochs. 

As explained in \Pa, PSF-fitting photometry was done with {\tt photutils} \citep{photutils}, basing the effective point-spread functions \citep[ePSFs;][]{Anderson_2000} on eight unsaturated and isolated stars in the field. 
%We fit the ePSF model to the data using a non-linear least-squares routine to infer the flux of the transient in each filter and epoch.
Both the Visit~1 image and the difference images have nonuniform backgrounds, but the gradients are much smaller in the difference images than in Visit~1.
For both image types, the uncertainties were estimated by injecting synthetic sources with flux equal to the measured flux of the transient at 150 nearby positions. The synthetic sources were then measured with the PSF-fitting photometry method, and the standard deviation of the  recovered fluxes was taken as the background uncertainty. 
Table~9 of \Pa\ gives the results.  As expected, the flux-density differences from Visit~1 have uncertainties about 3/4 of the uncertainty of the Visit~1 measurements.  In the four long-wave filters, W2 was not detected at all in Visit~1, but it was detected at 3--5$\sigma$ in the difference images.

We measure the photometry at the position of W2 in the F182M and F210M NIRCam images, using the PSF-fitting photometry method described above. The flux measurements are summarized in Table~\ref{tab_182_210_flux}.  There is no significant flux detected at the position of W2 in the F210M observation (six months after Visit 4) and the first F182M observation (11 months after Visit 4). These nondetections indicate that there is no underlying bright knot at the position of W2. A source is detected at the 3$\sigma$ level in the second F182M observation (18 months after Visit 4). 

{This 3$\sigma$ detection of a source at W2's position 18 months after the final PEARLS visit is not likely to be a reappearance of W2, since our kinematic model of W2 (see Section \ref{sec:orbit}) would not predict another caustic-crossing. This detection may instead be a separate microlensing event of a different star nearby W2, suggesting that W2 may exist inside a stellar cluster.}

\subsection{Stellar SED fitting}

{\Pa\ tested both single-star and binary-star fits to the eight-filter light curve from the PEARLS and CANUCS NIRCam visits. The fits were based on the {\tt pystellibs}\footnote{https://github.com/mfouesneau/pystellibs} with the BaSeL stellar library \citep{Lejeune_1998}.  Free parameters included values of the stellar effective temperatures $T_{\rm eff}$ and surface gravities $\log(g)$, stellar metallicity $Z_*/Z_{\sun}$, line-of-sight extinction due to dust $A_V$, and the ratio of total-to-selective extinction $R_V$.  All of the preceding values were held constant with time, but the flux-density normalizations of each star were free to vary independently for each visit.}

{The binary-star fits were compared to the single-star fits using the Bayesian Information Criterion \citep[BIC;][]{schwarz1978estimating}, which compares the goodness-of-fit of models with different numbers of free parameters (higher-complexity models are penalized). A difference in BIC values $\Delta {\rm BIC}>10$ indicates ``very strong" evidence that the model with the lower BIC value is preferred \citep{KassRaftery_1995}. In this case, the single-star model had nine free parameters and the binary-star model had 15 free parameters. The difference between the BIC values for the two models was $\Delta {\rm BIC}\equiv {\rm BIC_{ single} - BIC_{binary}}  = 26.7$, indicating ``very strong" evidence that the binary model was preferred over the single-star model.}  

Figure~\ref{fig:lightcurve} shows the resulting light curves in three example filters, and  Figure~\ref{fig:SED} gives the best-fitting spectral energy distributions (SEDs) in all four epochs.
The best-fitting temperatures of the two components were $\log(T_{\rm eff})=3.55^{+0.06}_{-0.08}$ and $\log(T_{\rm eff})=4.10^{+0.23}_{-0.13}$.  
Other fit parameters have large uncertainties --- meaning changing those parameters has a minor effect on the fit quality --- but they include supergiant surface gravities, reddening insignificant at the observed wavelengths, and  $Z_*\approx0.9Z_{\sun}$.

\subsection{Variability and Color Change}
W2's observed variability is not easily explained by intrinsic variability of its components.  While both RSGs and luminous blue variables (LBVs) can change brightness by more than a magnitude, luminous members of the respective classes exhibit such large changes only over timescales of years (e.g., \citealt{Yang_2011,Ren_2019,Spejcher_2025} and references therein), much longer than the timescales observed here.  
A better explanation for the observed variability is that the lensing magnification is changing over time.  The rapid changes come from temporary boosts by microlensing as an intracluster star or other compact object changes its  alignment with the lensed source. The dramatic color changes (colors becoming bluer by $\sim 200$\%:  Figure~\ref{fig:color_curve}) can be explained by the binary star's orbital motions across the microlensing caustic. The blue, hot component of the SED becomes brighter with time, and continuing to do so over $\sim 60$ rest-frame days is inconsistent with astrophysical explosions and eruptions.

%We show the 126-day light-curve of the source in the F115W, F200W, and F356W filters in Figure \ref{fig:lightcurve}. The

%dramatic difference in the shape of the light-curves between the filters indicates that the system consists of a red component and a blue component. The blue component is detected in all four epochs and gets brighter over time, suggesting that its magnification is increasing as it approaches the critical curve. The red component is undetected in Visit 1, peaks in brightness in Visit 2, and then gets dimmer in Visits 3 and 4, indicating that it is on the opposite side of the critical curve in Visit 1 (where magnification is zero).

%Assuming a maximum possible magnification of $\mu\approx 20.000$, a main-sequence star with $\log(T_{\rm eff}/K)=3.55$  would not be bright enough to be detected in the JWST NIRCam data. Therefore, the cooler star is likely an extremely luminous evolved post-main sequence star. The maximum magnification is inversely proportional to the square root of the source's radius, so an evolved red star with $R\gtrsim100R_\odot$ requires a luminosity in the supergiant regime in order to be detectable.  

\section{Approximate source-plane size constraints}\label{sec:size}

The observed light curve of the lensed source shows significant variability over the 126-day baseline, suggesting 
%that the magnification is varying with time due to 
microlensing by an M0416 intracluster star. 
%If the source was magnified only by the M0416 cluster, we would not observe this time-variable magnification. 
This microlensing constrains the projected source-plane size of the lensed source, $R_{s\perp}$ \citep[e.g.,][]{Vovk_2016}. 
%
%Microlensing can magnify the flux of a background source if 
The size of the source must be comparable to or smaller than the Einstein radius of the microlens at the source's redshift, 
%The Einstein radius $\theta_E$ is given by
\begin{equation}
    \theta_E = \sqrt{\frac{4GM}{c^2}\frac{D_d D_{ds}}{D_s}} \quad,
\end{equation}
where $G$ is the gravitational constant, $M$ is the mass of the microlens, $c$ is the speed of light, and $D_d$, $D_s$, and $D_{ds}$ are the angular diameter distances from (respectively) the observer to the lens, the observer to the source, and the lens to the source. For a microlens mass of 1.0~$M_\sun$, the Einstein radius for an isolated microlens is $\theta_E\approx1700$~AU. 

Obtaining a source-plane size estimate from the Einstein radius of the microlens requires accounting for the effect of the galaxy-cluster (M0416) lens. For a given microlens Einstein radius $\theta_E$ and radial and tangential microlens magnifications $\mu_r$ and $\mu_t$ (respectively), the maximum size of the caustic $\beta_{\rm max}$ is given by
\begin{equation}
    \beta_{\rm max} \approx \frac{1}{\mu_r}\sqrt{\frac{|\mu_t|}{8}}\theta_E
\end{equation}
\citep{Oguri_2018}. For a {\tt GLAFIC} (V4) lens model of M0416 \citep{oguri10, kawamataoguriishigaki16, kawamataishigakishimasaku18}, the magnifications at the position of the source are $\mu_r\approx1.67$ and $\mu_t\approx192$. These values yield a maximum projected source-plane size $R_{s\perp}\lesssim5000$~AU.
The most compact stellar clusters have effective radii of order $0.1$~pc (20,000 AU) \citep[e.g.,][]{Ryon_2015}, so this constraint suggests that the source could be a very small group of stars or a single stellar system, but not a large stellar cluster.

% The amplitude of the change in magnification yields an additional constraint on $R_{s\perp}$. At the moment of maximum magnification (where the microlens is directly aligned with the source), an order of magnitude estimate for the microlensing magnification is given by,

% \begin{equation}
%     \mu_{\rm micro} \approx \sqrt{\frac{R_E}{R_{S\perp}}}
% \end{equation}

% \noindent from \cite{Chang_1984}. The maximum observed change in flux is in the F115W filter, where the flux increases by a factor of $\sim$3.4 between Epoch 1 and Epoch 4 (see Figure \ref{fig:lightcurve}. This allows us to estimate $\mu_{\rm micro}\approx3.4$, giving an order of magnitude estimate for the projected size of the lensed system of $R_{S\perp}\approx150$~AU. 

% \begin{equation}
%     \vec{V}_{ T,s}=\frac{R_{s\perp}}{\Delta t}
% \end{equation}

A more stringent source-plane size constraint comes from the timescale of the microlensing event $\Delta t$. The duration of the microlensing event is determined by the projected transverse velocity of the microlensing object with respect to the source and the projected size of the lensed source \citep{Kayser_1986}. The velocity of the microcaustic is given by
%\begin{equation}
%    \Delta t =v_\perp R_{s\perp}
%\end{equation}
\begin{equation}
\vec{V}_{T, s} = \frac{\vec{v}_s}{1 + z_s} + \frac{\vec{v}_{\text{obs}}}{(1 + z_d)}\frac{D_{ds}}{D_d } - \frac{\vec{v}_T}{(1 + z_d)}\frac{D_s}{D_d},
\end{equation}
where 
%$D_{ds}$, $D_{d}$, and $D_{s}$ are the angular diameter distances, 
$\vec{v}_{\text{obs}}$ is the velocity of the observer, $\vec{v}_T$ is the velocity of the lens, and $\vec{v}_s$ is the velocity of the source. 
%As explored by \citet{Windhorst_2018},

Because the source was detected in all four epochs, the timescale $\Delta t$ must be comparable to or longer than the 126-day baseline of the observations. {The size constraint scales linearly with the assumed event duration. To be conservative, we assume the maximum total duration of the microlensing event is approximately twice as long as the 126-day baseline, or $\Delta t\approx250$~days.}  For a transverse velocity approximately equal to the velocity of the Sun with respect to the cosmic microwave background reference frame, $v_{\perp}\approx300$ km s$^{-1}$ \citep{Kogut_1993}, the source-plane size would be $R_{s\perp}\lesssim45$~AU\null. {For an even longer assumed event duration of 500 days, the source plane size constraint would be $R_{s\perp}\lesssim90$~AU\null. Section~\ref{sec:orbit} suggests that the transverse velocity may be $\sim 50$ km s$^{-1}$, corresponding to a proportionally smaller projected source size or longer event duration.}

Finally, the change in the color of the SED (Figure~\ref{fig:color_curve}) implies that the color temperature of the source must vary on a physical scale smaller than  126~days/50~km~s$^{-1}$ $\approx 135$~AU (22.5~AU for 300 km s$^{-1}$). This change in color over a small physical scale can be explained by the orbital dynamics of a binary star system causing the magnification of each star to vary differently with time (as discussed in Section \ref{sec:orbit}). 

\section{Binary configuration search}\label{sec:posydon}
\begin{figure*}[ht]
    \centering
    \includegraphics[width=\linewidth]{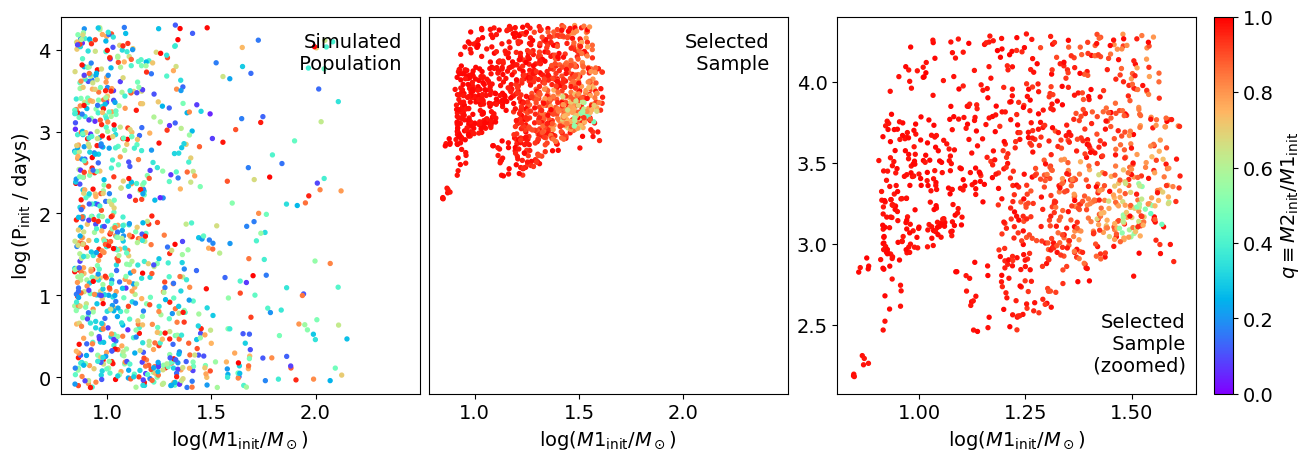}
    \caption{{\it Left:} Initial masses and periods of all 100,000 binary systems in the simulated population. {\it Middle:} Initial masses and periods for the 1001 systems that can reproduce the observed transient SED with $\mu<10,000$. {\it Right:} Same as middle, but zoomed in on the selected population. The color of each point indicates the configuration's initial mass ratio $q$.}
    \label{fig:population}
\end{figure*}

\subsection{Binary Simulations}\label{subsec:binary_selection}

{To search for possible binary stellar systems that can reproduce W2's observed SED, we use the binary stellar population synthesis code {\tt POSYDON v1.0} \footnote{\href{https://github.com/POSYDON-code/POSYDON/commits/v1.0.5/}{{\tt POSYDON} github source code}} \citep{Fragos_2023}. {\tt POSYDON} employs the stellar-structure code Modules for Experiments in Stellar Astrophysics \citep[MESA;][]{Paxton2011,Paxton2013,Paxton_2015,Paxton2018,Paxton2019} to create grids of binary stellar-evolution models\footnote{{\tt POSYDON} Data Release 1 grids are archived at \href{https://zenodo.org/records/6655751}{https://zenodo.org/records/6655751}}, with separate grids created for distinct phases of binary evolution.
The simulated stars generated by {\tt POSYDON v1.0} have stellar metallicity fixed to the solar value, while the nebular oxygen abundance in the Warhol arc is slightly subsolar, $\log(Z_{\rm neb}/Z_\odot)=-0.24\pm0.08$ (\Pa). We do not expect this small difference to significantly affect the results.}
%We used the binary stellar population synthesis code {\tt POSYDON v1.0} \citep{Fragos_2023} to search for possible binary stellar systems that can reproduce W2's observed SED\null. {\tt POSYDON} employs the stellar-structure code Modules for Experiments in Stellar Astrophysics \citep[MESA;][]{Paxton2011,Paxton2013,Paxton_2015,Paxton2018,Paxton2019,Jermyn2023} to create grids of binary stellar-evolution models with separate grids created for distinct phases of binary evolution. 

\begin{figure*}[ht]
    \centering
    \includegraphics[width=0.7\linewidth]{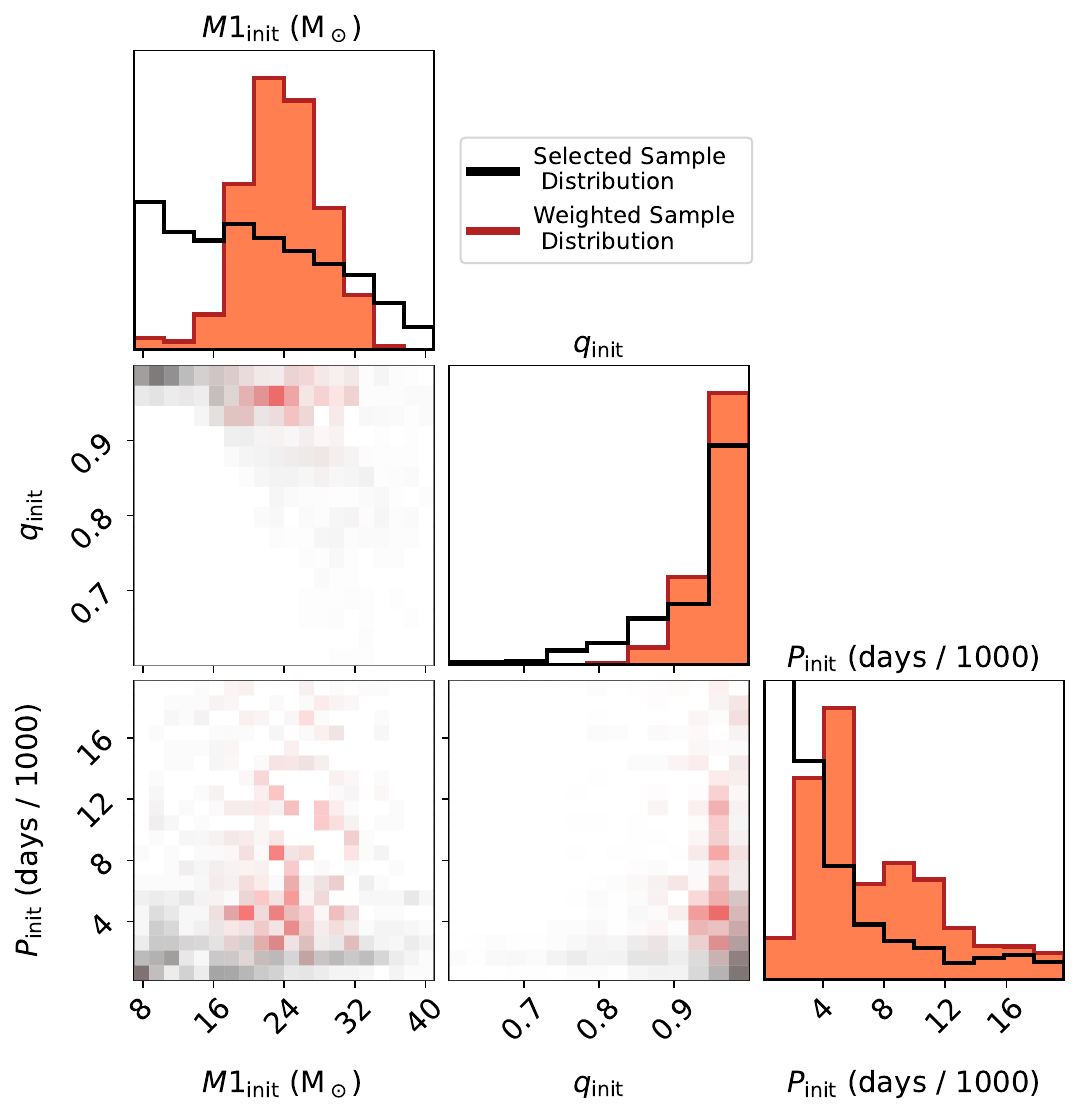}
    \caption{Distribution of initial mass, initial mass ratio $q$, and initial orbital period for the final sample of binary configurations that can reproduce the observed SED of the lensed binary. The black histograms show the distributions for the unweighted selected sample, and the red histograms show the selected sample weighted by time and magnification.}
    \label{fig:hists}
\end{figure*}

We used  {\tt POSYDON} to generate a synthetic population of 100,000 binary stellar systems. The initial mass of Star~1 ($M1$) was stochastically drawn from the \cite{Kroupa_2001} initial mass function (IMF), initial mass ratios $q\equiv M2/M1$ were drawn from a flat distribution spanning the range 0.1--1.0, and initial orbital periods ($P$) were drawn from the log-space power law described by \cite{Sana_2013}. {The minimum period was set such that zero-age main-sequence systems did not undergo Roche-lobe overflow, and the maximum period was set to 20,000 days.} For each initial binary configuration, {\tt POSYDON} used nearest-neighbor interpolation to the closest MESA binary grid, tracking its (downsampled) evolution from the zero-age main-sequence (ZAMS) state to the final state as a disrupted binary, binary  merger, or until the formation of a compact object (white dwarf, neutron star, or black hole). 

{To select binary configurations that could reproduce W2's SED, we apply two selection criteria to the simulated population of 100,000 binaries. First, we select those systems where the temperatures of the two stars at any epoch of evolution are within 2$\sigma$ of W2's stellar temperatures (\Pa). Next, we require that the stellar luminosities are high enough to reproduce W2's observed fluxes in all filters and epochs, assuming a maximum allowed magnification of $\mu=10,000$ for each star. To compute the necessary magnifications for each configuration, we generate a model spectrum of each star using {\tt pystellibs} with the BaSeL library of stellar atmospheres \citep{Lejeune_1998}, and we apply {\tt pysynphot}\footnote{https://github.com/spacetelescope/pysynphot} to generate each star's \textit{unmagnified} synthetic NIRCam photometry. We then used chi-squared minimization to find the best-fitting magnifications for each epoch, allowing the magnification of each star in the binary to vary independently. Figure~\ref{fig:SED} shows an example of a single model system fit to the four epochs.}

{Our selected sample of simulated binary configurations which satisfy both the temperature and magnification criteria comprises 1001 systems. Figure~\ref{fig:population} shows the initial masses and  periods of the systems in the full simulated population and in the selected sample, and the parameter distributions in the selected sample are shown in Figure \ref{fig:hists}.}

Viable initial masses of Star~1 range from $M1_{\rm init}=8$~\Msol\ to 45~\Msol. The lower mass limit arises because less-massive stars cannot reach $\log(T_{\rm eff}/{\rm K})=4.10$. {The upper mass limit arises because stars with $M_{\rm init}\gtrsim40$~\Msol\ evolve into luminous blue Wolf--Rayet stars rather than cool RSGs, so Star 1 never becomes an RSG above this mass limit. Stars with $M_{\rm init}\gtrsim40$~\Msol\ would not have long enough lifetimes to support the evolution of a companion star into an RSG, so Star 2 also never becomes an RSG above this mass limit.} The systems in the final sample tend to have $q$ close to unity because the luminosity of the two sources must be comparable for both stars to be detected. 

The initial orbital periods in the selected sample range from $P_{\rm init}=180$~days up to the maximum period generated by {\tt POSYDON}. Short-period systems are ruled out because the initially larger star undergoes too much mass loss during its main-sequence lifetime and becomes a blue, helium-stripped star rather than an RSG\null.  Both interacting and noninteracting systems can reproduce the  W2 observations, but 59\% of the selected binary configurations experience some amount of mass transfer when the initially more massive star expands into its RSG phase and fills its Roche lobe.

\subsection{Rarity of Selected Systems}

{To estimate how rare the binary configurations in the selected sample are, we compute the total underlying stellar mass of the entire simulated population of 100,000 binary configurations.}

{{\tt POSYDON} estimates the total stellar mass of the underlying stellar population (i.e., including single stars along with the simulated binaries) assuming a binary fraction of 0.7. For our our full 100,000-star simulated population, $M_* =10^{7.20}$~\Msol. After correcting for magnification, the total stellar mass of the Warhol arc is of order $10^6$~\Msol\ \citep{Palencia_2025}, so the
1001 selected binary configurations scale to 63 in Warhol's stellar mass. If the stellar population simulated by {\tt POSYDON} is representative of the stellar population in the Warhol arc, we expect that approximately 63 binaries in the galaxy would satisfy the selection conditions described in Section \ref{subsec:binary_selection}.}

{With a maximum possible magnification $\mu=10,000$, the simulated sample produces 29318 systems in which both stars would be detectable at 5$\sigma$ in at least two NIRCam filters (i.e. 29318 simulated systems satisfy the magnification condition but do not necessarily satisfy the temperature condition). Therefore, our selected sample of 1001 binary systems represents approximately 3\% of the \textit{detectable} stellar population.}

\subsection{The Likeliest Binary Systems}

\begin{figure}[ht]
    \centering
    \includegraphics[width=\linewidth]{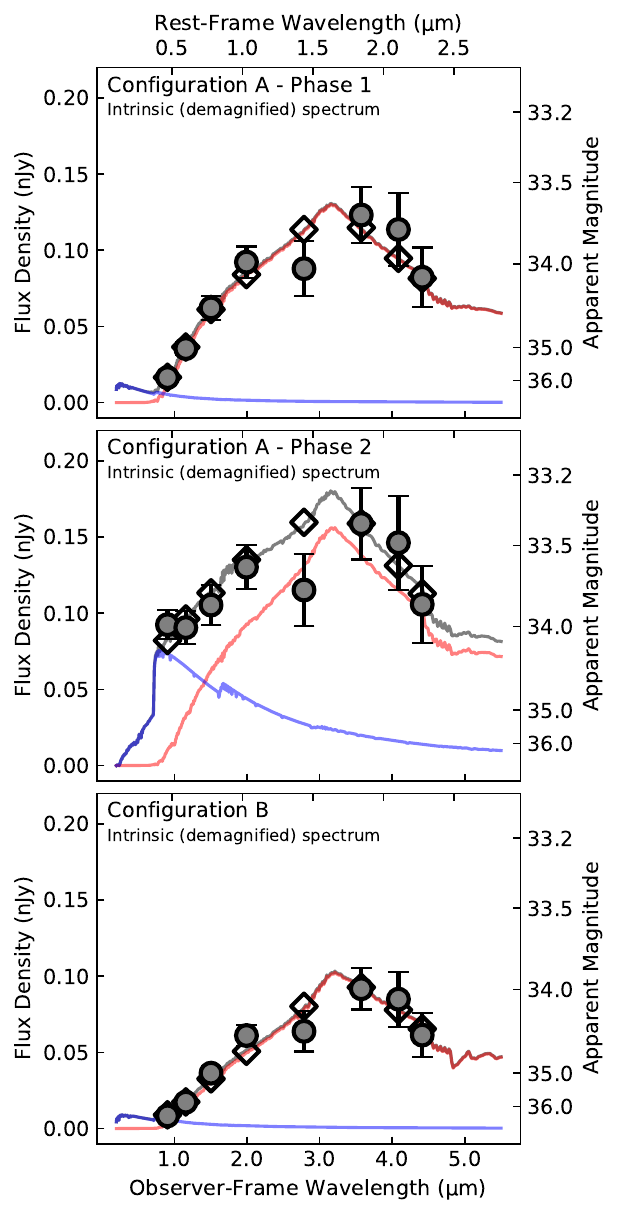}
    \caption{Intrinsic spectra of two selected simulated systems, ``Configuration A" and ``Configuration B". The top two panels show the two distinct phases of the Configuration A system. Phase one is when Star 1 is the RSG, and Phase two is when Star 2 is the RSG. Plotting symbols are the same as in Figure \ref{fig:SED}, except the observed photometry has been de-magnified according to the magnification values inferred for each configuration.}
    \label{fig:native_SEDs}
\end{figure}

While the 1001 binary systems in the selected sample are all possible configurations for W2, they are not equally likely. The duration of the evolutionary phase during which the stars would have the necessary temperatures and luminosities varies from $< 1000$~yr for some systems to $> 100,000$~yr for others.

The probability of observing a given binary configuration is directly proportional to the duration. 
Magnification also affects which systems are most likely to be observed.  Lower magnifications are more common than higher ones, and therefore systems with higher luminosity are more likely to be observed. (The initial Kroupa IMF in the sample construction accounts for such systems being less common.)

The probability of observing a lensing event with a given magnification depends on the lens' properties.  The probability density function (PDF) can be derived from the M0416's stellar-mass surface density as traced by the intracluster light $\Sigma_{\rm ICL}$, the magnification from the cluster $\mu_{\rm macro}$, and the critical density for lensing $\Sigma_{\rm crit}$.
In the region near the Warhol arc, $\Sigma_{\rm ICL}\approx59$~\Msol~pc$^{-2}$ \citep{Kaurov_2019}, and $\mu_{\rm macro}$ and $\Sigma_{\rm crit}$ come from the  {\tt GLAFIC} lens model \citep[v4;][]{oguri10, kawamataoguriishigaki16, kawamataishigakishimasaku18}.  For each binary configuration in the selected sample, we computed the probability of the magnification needed to reproduce W2's SED from the \cite{Palencia_2024} PDF.
\begin{figure*}
\centering
%\begin{subfigure}{\textwidth}
    \includegraphics[width=\textwidth]{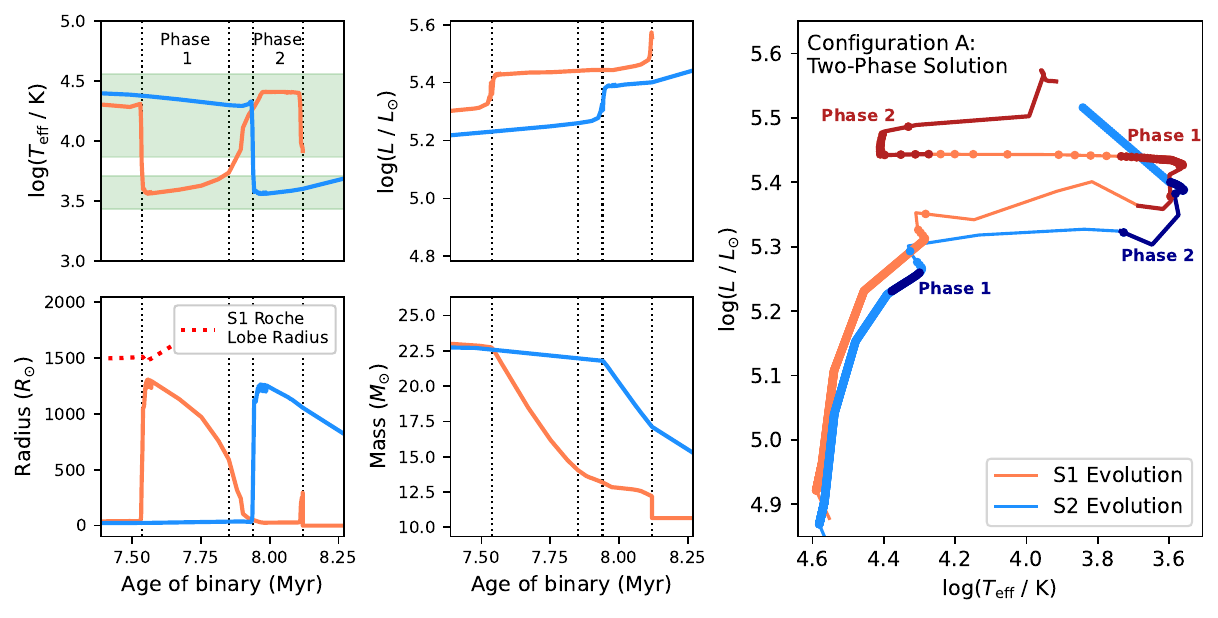}
%\end{subfigure}
%\end{figure*}
%
%\begin{figure*}
    \includegraphics[width=\textwidth]{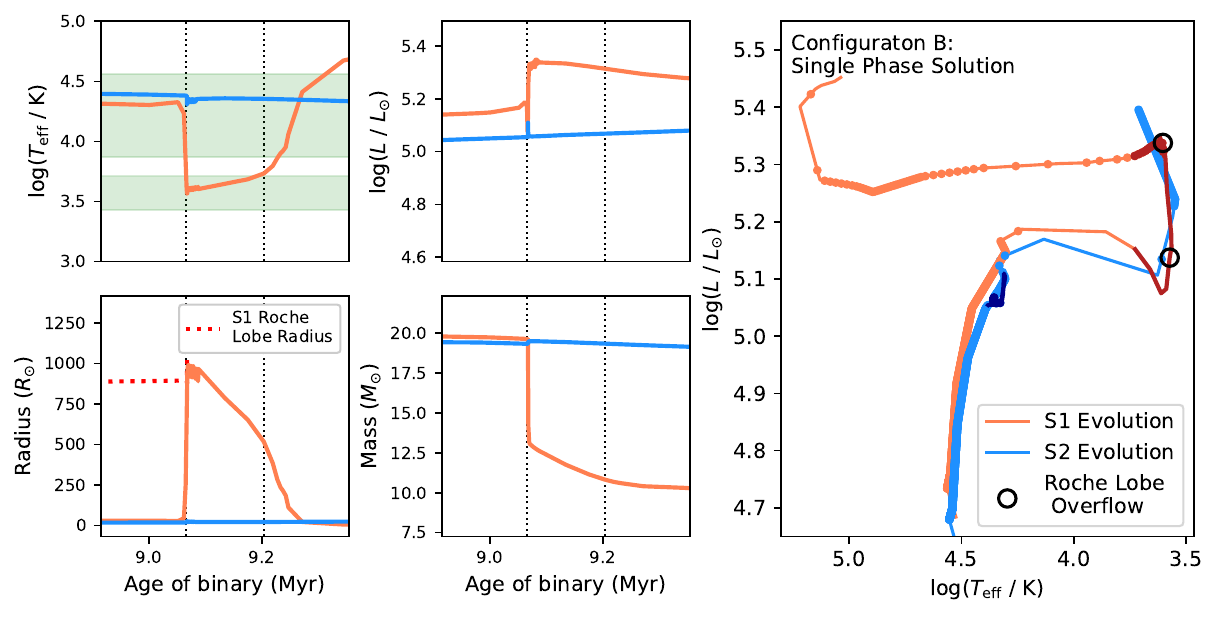}
        \caption{Evolutionary history of two possible configurations for W2.  Configuration~A in the top five panels is a noninteracting binary with a two-phase solution, while Configuraton~B in the bottom five panels is a single-phase solution. The small panels show stellar parameters as a function of age with blue lines for the initially more-massive star and orange lines for the less-massive one. The green-shaded regions in the top-left plot for each configuration indicate the accepted temperature ranges. The gray dashed lines bound the age ranges for which these configurations can reproduce W2's observed SED\null. The large panels at right show H--R diagrams for the two configurations with the same color coding and bold regions indicating the acceptable age ranges.  The points in these panels indicate equal  10,000~year intervals, and the black circles in the lower diagram indicate the start and end of Roche-lobe overflow. 
         Configuration~A has $M1_{\rm init}=25.5$~\Msol, $q_{\rm init}=0.95$, and $P_{\rm init}=4024$ days. Configuration~B has $M1_{\rm init}=21.6$~\Msol, $q_{\rm init}=0.95$, and $P_{\rm init}=2043$ days.}
    \label{fig:evol}
    \label{fig:evolA}
    \label{fig:evolB}
\end{figure*}

The final weight of each binary configuration in the selected sample is the product of duration of the required evolutionary phase and the magnification probability. Figure \ref{fig:hists} shows the parameter distributions of both the weighted and unweighted samples. {The most likely M1$_{\rm init} = 23.6^{+5.3}_{-4.3}$ M$_\odot$. This skews slightly higher than the unweighted value because higher-mass stars need less magnification. The weighted distribution of initial orbital periods spans a wide range, P$_{\rm init} = 5528^{+6372}_{-2145}$~days. This corresponds to orbital separations of 5--50~AU, consistent with the size constraints in Section~\ref{sec:size}. The most likely initial mass ratios are near unity, q$_{\rm init} = 0.96^{+0.02}_{-0.03}$. }

{For 37\% of the binary configurations in the weighted sample, there are two distinct phases during the binary's evolution during which the system can reproduce the observed SED of the lensed binary. The first phase happens when the temperature of Star~1 (initially larger mass) drops and it becomes an RSG, while Star~2 (initially lower mass) remains hot. The second phase happens when the temperature of Star~1 increases again shortly before it undergoes core collapse, and by coincidence the temperature of Star~2 drops as it becomes an RSG\null. If W2 represents the second phase of a two-phase configuration, we would be observing a massive binary seen immediately before the RSG explodes as a Type~II supernova.}

{To determine the probability of W2 being observed in the second phase of a two-phase configuration, we repeat the weighting by duration and magnification on each phase separately. The first phase has a longer duration than the second phase, but the stellar luminosities are higher during than second phase, resulting in lower magnification factors during the second phase. Therefore, the weights for the first and second phase are approximately equal (54\% and 46\%, respectively).  Since the two-phase systems represent 37\% of the weighted sample, the probability of observing W2 in the second phase of a two-phase configuration is 17\%.}

{Interacting binaries are less likely in the weighted sample: 20\% of the configurations in the weighted sample experience some amount of mass transfer during their evolutionary histories. This fraction is significantly lower than the unweighted sample's 59\%, suggesting that noninteracting systems with longer initial orbital periods last longer. The  two-phase configurations and single-phase configurations both consist of approximately 20\% interacting binaries.}

{Figure \ref{fig:native_SEDs} shows the intrinsic (demagnified) simulated spectra of two likely binary configurations from the selected sample. ``Configuration A" is an example of a ``two-phase'' solution, and we show the intrinsic spectra from each phase in separate panels. ``Configuration B" is an example of a single phase solution.  Configuration A has initial mass $M1_{\rm init}=25.5$~\Msol, $q_{\rm init}=0.95$, and $P_{\rm init}=4024$ days. Configuration B has $M1_{\rm init}=21.6$~\Msol, $q_{\rm init}=0.95$, and $P_{\rm init}=2043$ days. Figure \ref{fig:evol} shows the evolutionary history of both configurations.}

\subsection{Comparison to known RSG binaries}
{To assess the physical validity of the simulations, we search for known RSG binary systems whose parameters are comparable to the most likely parameters inferred for W2. A small number of known RSG binaries have sufficient observational constraints to characterize their orbital parameters. Statistical analyses suggest that the distribution of orbital periods of high-mass binary systems peaks at short periods ($P<10$ days) and follows a power-law distribution out to $>10^4$ AU \citep{Duchene_2013}.} 

{We identify two well-constrained RSG binary systems in the Milky Way whose parameters are comparable to the most likely parameters for W2. VV Cephei is an eclipsing binary consisting of an RSG and a blue companion star, and its orbital parameters have been constrained via photometric and spectroscopic monitoring over multiple orbits \citep{Wright_1977,Thomas_2015,Pollmann_2020,Pollmann_2023}. The RSG in VV Cephei has a mass $\approx20$ M$_\odot$, the mass ratio of the companion to the RSG is close to unity, and the orbital period of the system is 7431 days \citep{PatrickNeguerela2024}.}

{A second comparable system to W2 is KQ Puppis, an RSG with a hot companion star in a highly eccentric orbit \citep{Cowley_1965,Jaschek_1963,Rossi_1992}. The RSG in KQ Puppis has a mass $\approx 13-20$ M$_\odot$, the mass ratio of the companion to the RSG is close to unity, and the system's orbital period is 9752 days \citep{PatrickNeguerela2024}.}

%We show the evolutionary history of two likely binary configurations in Figures \ref{fig:evolA} and \ref{fig:evolB}. Configuration A is an example of a non-interacting binary with a ``two-phase'' solution, and Configuration B is an example of an interacting binary with a single phase solution.  Configuration A has initial mass $M1_{\rm init}=25.5$~\Msol, $q_{\rm init}=0.95$, and $P_{\rm init}=4024$ days. Configuration B has $M1_{\rm init}=21.6$~\Msol, $q_{\rm init}=0.95$, and $P_{\rm init}=2043$ days.

\section{Light-Curve Modeling}\label{sec:orbit}

If the orbital velocities of the stars in a microlensed binary system are comparable to or larger than the transverse velocity of the microlensing caustic, the light curves will show unique patterns as the each star approaches or recedes from the microlensing caustic with a different velocity compared to its companion \citep{Zheng_2025}. As shown in Figure \ref{fig:lightcurve}, the light curves for different NIRCam filters differ dramatically. If both stars were approaching the microlensing caustic at the same velocity, their magnification factors would increase at the same rate, and the color (Figure \ref{fig:color_curve}) would not change.  The observed color change therefore implies that the orbital velocities of the stars dominate over the transverse velocity of the microlens.

For an extended source with radius $R$, the magnification as a function of projected distance $l$ from the critical curve is given by
\begin{equation}
    \mu = \frac{2A_0}{\pi\sqrt{R}} \int_a^b \sqrt{\frac{y(2-y)}{y+y_0}} dy\quad,
    \label{eqn:miralda}
\end{equation}
where $y_0=(l/R) -1$, and the integral is computed over the bounds [$0$,$2$] for $y_0>0$ or [$-y_0$,$2$] for $y_0<0$ \citep{Miralda_1991}. Here, $l$ and $R$ are given in units of AU, and the normalization factor $A_0$ is in AU$^{0.5}$ to make the magnification dimensionless.

To investigate whether the binary-system hypothesis is plausible and estimate the range of orbital parameters,
we repeated the chi-squared minimization described in Section~\ref{sec:posydon} but fit for the system's kinematics 
%with respect to the critical curve 
rather than allowing the magnifications to vary independently for each star at each epoch. The model's free parameters were the normalization of the magnification $A_0$, the system's perpendicular velocity with respect to the microlensing caustic $v_{\perp}$, the initial distance (setting time $t=0$ at Visit~1) from the binary system's center of mass to the microlensing caustic $l_0$, the orbital eccentricity $\epsilon$, the angular orientation of the binary's ellipse with respect to the microlensing caustic $\omega$, and the  angular position at Visit~1 of Star~1 along its orbital path $\nu_0$. In principle, the inclination angle of the ellipse along both axes should be two more free parameters, but for simplicity they were fixed at zero, (face-on). Inclined viewing angles would make the magnification variation less dramatic, and an edge-on orbit would be unable to reproduce the observed light curve. The binary's orbital period and separation, the radius of each star, and the orbital velocities of each star were fixed to the  {\tt POSYDON} model values. {\tt POSYDON} assumes that all orbits are circular throughout the systems' evolution, so the eccentricity is a free parameter in our kinematic model. 

\begin{table}[ht]
    \centering
    \ra{1.5}
    \caption{\textbf{Light-Curve Modeling}}
    \begin{tabular}{ccc}
    \toprule\toprule
     & \multicolumn{2}{c}{Best-Fitting Values}  \\
    Free Parameter & Configuration A & Configuration B\\

    \midrule
        $A_0$ & 3373 AU$^{-1/2}$ & 4022 AU$^{-1/2}$   \\
         $v_\perp$& $-46.3$ km s$^{-1}$ & $-57.3$ km s$^{-1}$\\
         $l_0$& $-0.12$ AU & 1.07 AU\\
         $\epsilon$ & 0.79 & 0.59\\
         $\omega$ & $159^{^{\rm o}}$ &  $184^{^{\rm o}}$ \\
         $\nu_0$ & $242^{^{\rm o}}$ & $233^{^{\rm o}}$ \\
    
    \bottomrule
    \end{tabular}
    
    \footnotesize{{\bf Notes.} Best-fitting values for the free binary-orbit parameters for the two binary configurations shown in Figure~\ref{fig:evol}.}
    \label{tab:orbit_Fit}
\end{table}

Each set of model parameters gives the stars' positions as a function of time (corrected for time dilation), and the projected distance from the caustic $l$ then gives the magnification through Equation~\ref{eqn:miralda}.
%is computed by combining the velocity of the binary's center of mass with respect to the critical curve and the star's orbital velocity. The binary's motion is corrected for redshift time dilation. The magnification of each star at a time $t$ is then given by 
%
Table~\ref{tab:orbit_Fit} gives the
best-fitting kinematic parameters for the same two likely {\tt POSYDON} configurations as are shown in Figure~\ref{fig:evol}. 

%Configuration A is an example of a non-interacting system, with $M1_{\rm init}=25.5M_{\odot}$, $q_{\rm init}=0.95$, and $P_{\rm init}=4024$ days. Configuration B is an example of an interacting system, with $M1_{\rm init}=21.6M_{\odot}$, $q_{\rm init}=0.94$, and $P_{\rm init}=2043$ days.

\begin{figure*}
    \centering
    \includegraphics[]{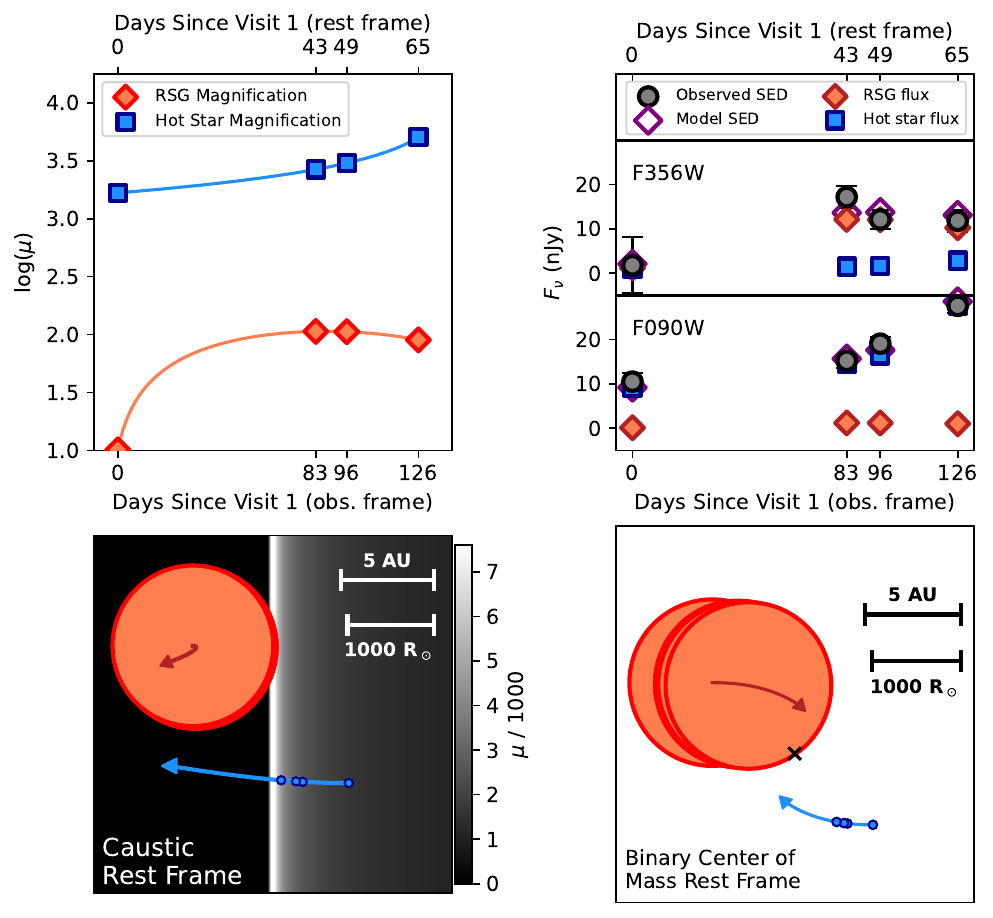}
    \caption{\small Best-fitting orbit for Configuration A\null. The top-left panel shows magnification as a function of time for each star. The top-right panel gives the model and observed light curves in the F356W and F090W filters. The bottom-left panel illustrates each star's path with respect to the microlensing caustic in the caustic's rest frame, and the shaded grayscale shows the magnification for an object with radius equal to the radius of the blue star (Table~\ref{tab:orbit_Fit}).  The four small circles show the model position of the blue star at each of the four epochs starting with Visit~1 at the right end of the line.  The model has the blue star approaching the critical curve on the magnified side, while the majority of the RSG is on the opposite side of the critical curve and therefore has a much lower magnification. The bottom-right panel shows the binary's orbit in the center-of-mass rest frame with the black ``x'' marking the center of mass. In both of the bottom panels, the large red circles show the model position of the RSG at the time of each NIRCam visit, and the red line with an arrow represents the RSG's orbital path.  (The RSG's motion is much smaller than its diameter and almost negligible with respect to the caustic as the orbital motion nearly cancels the center-of-mass' transverse motion.) The blue line with arrow shows the blue star's orbital path. The sizes of the red and blue circles and their separations are all on the scale indicated, except the radii of the blue circles are increased by a factor of 3 for visual clarity.}
    \label{fig:orbitA}
\end{figure*}

\begin{figure*}
    \centering
    \includegraphics[]{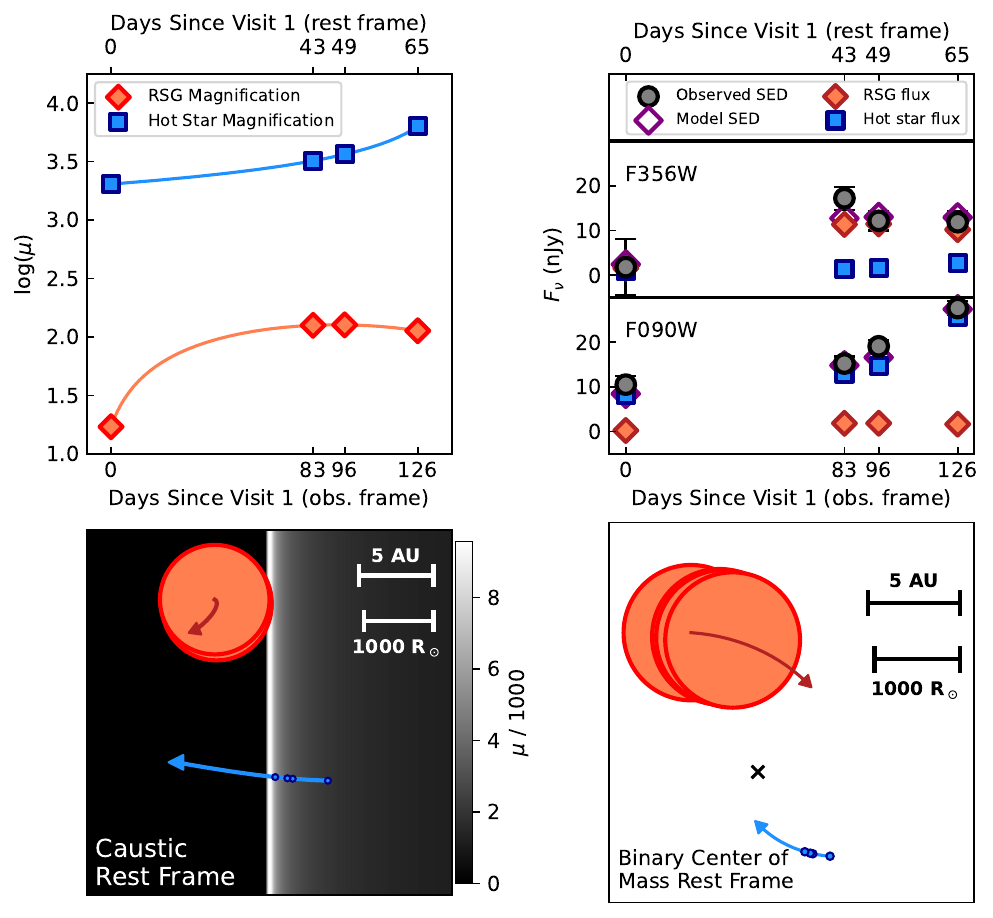}
    \caption{\small Same as Figure \ref{fig:orbitA}, but for binary Configuration B.}
    \label{fig:orbitB}
\end{figure*}

The kinematic models can reproduce the observed light curves of the lensed binary for both configurations in all filters with $\chi^2_r \approx1.0$. Both configurations favor an eccentric orbit ($\epsilon=0.79$ for Configuration~A, $\epsilon=0.59$ for Configuration~B), an initial center-of-mass position within $\sim 1.0$~AU of the microlensing caustic, and a projected transverse velocity with respect to the critical curve $v_\perp\approx-50$~km~s$^{-1}$. The best-fitting orbits and the resulting light curves are shown in Figures~\ref{fig:orbitA} and~\ref{fig:orbitB}.

\section{Alternative Scenarios}
Some alternative scenarios to a binary stellar system and their abilities to reproduce the observed SEDs and light curves are described below.

{\textbf{Alternative 1, two independent stars:}} the source-plane size constraints in Section~\ref{sec:size} refer to the {\em projected} size of the source. 
%We consider the possibility that the observed source consists of 
Two independent stars that are close in projection but distant from one another in the line of sight or are magnified by a different caustic or region of the same caustic could in principle reproduce the light curve.  In effect, it would be a coincidence that two stars with the right evolutionary states and motions (and therefore magnifications as a function of time) happen to fall along the line of sight. 
%{\color{red} insert probability of two events at same location. independent measurements of the centroid of the source in each epoch agree within 0.021 arcsec, what is probability of observing multiple events in that small radius?}

{\textbf{Alternative 2, a single nonvariable star:}} a single massive OB star or RSG could be luminous enough to reproduce the observed fluxes even with $\mu<10,000$ and compact enough to experience the observed magnification-variation due to microlensing. However, no single-star SED  fits the observations ($\Delta {\rm BIC} = 26.7$), and a single nonvariable star would not change color over time, contrary to observations (Figure~\ref{fig:color_curve}). 

{\textbf{Alternative 3, a single variable RSG:}} an individual RSG with time-variable luminosity and temperature, such as a Mira variable, could be luminous  enough to be detected with $\mu<10,000$, and it would change in color over time as the temperature varies. Mira variable stars are evolved red giants with radii $\sim400$~R$_\sun$ which pulsate and vary by up to $\sim30\%$ in temperature and a factor of $\sim 2$ in luminosity over a cycle of $\sim 1$~yr  \citep[e.g.,][]{Reid_2002}. While a Mira's color could change over a similar timescale to the observed 126-day light curve, a temperature change of $\sim30\%$ would not be enough to reproduce the observed color change. Even worse, no single star could reproduce the two-component SED in any one epoch. For Miras in particular, the maximum temperature would never be hot enough to reproduce the observed blue component of the SED. 

{\textbf{Alternative 4, a single variable OB-type star:}} LBVs are extremely luminous, hot stars that exhibit significant luminosity and temperature variations (\citealt{Weis_2020} provide a recent review). An LBV star would be luminous enough to be detected and compact enough to experience microlensing. The temperature fluctuations of LBVs can be as large as $0.6$~dex, so an LBV would exhibit a strong change in color over time. However, the typical timescales of large-scale LBV temperature variations are on the order of a decade, and shorter timescale variations are much smaller \citep[e.g.,][]{Sterken_2003}. Additionally, no single star can reproduce the SED at any epoch.  For LBVs in particular, the typical minimum temperature reached during coolest state is $\sim 6000$--8000~K, far too hot to reproduce the red component of the observed SED ($\sim 3600$~K). Therefore, an LBV star alone cannot reproduce the observations.

Type Be supergiants are evolved massive stars surrounded by large amounts of circumstellar gas and dust \citep[e.g.,][]{Aret_2012,Maravelias_2023}. A red component in the SED might arise from hot dust in the circumstellar material, but the dust temperature would have to be below the sublimation temperature of dust, and the optical depth would have to be tuned to $\sim 0.5$ to allow both hot and cold components to be visible. 
%However, the luminosity of the red component would not be bright enough to reproduce the observed fluxes in the long-wavelength filters, assuming a maximum magnification $\mu\lesssim20,000$. Additionally, 
Even with such fine tuning, the system would be small enough that  the cool and hot components would have identical magnifications, so this scenario could not reproduce the observed change in color.

{\textbf{Alternative 5, a compact star cluster:}} a compact stellar cluster positioned very close to the lensing caustic could be luminous enough to be detected, and a cluster containing both hot and cool stars could reproduce a multicomponent SED\null. However, the factor of $\sim 3.4$ increase in flux over 126-days confirms microlensing by an intracluster star. This constrains the projected source size to $\lesssim 90$~AU (Section \ref{sec:size}), ruling out a star cluster.

{\textbf{Alternative 6, a black-hole accretion disk:}} a thin accretion disk surrounding a black hole could be luminous enough to reproduce the observed fluxes and compact enough to display magnification variation due to microlensing. Because the temperature of an accretion disk varies with radius \citep[$T_{\rm eff}\propto r^{-3/4}$;][]{Blackburne_2011}, a microcaustic moving across an accretion disk could produce an observed change in color over time as the position of the highly magnified region changes. However, a microlensed accretion disk surrounding a stellar-mass black hole would emit mostly in rest-frame X-rays, which would not be detectable in the JWST NIRCam filters at Warhol's redshift. (\citealt{Windhorst_2018} investigated the detectability of these systems at $z\gtrsim7$.) Additionally, while this scenario could produce an observed change in color over time, it would not reproduce the observed two-component SED in any one epoch. 

\section{Discussion and conclusions}\label{sec:conclusion}
The most likely explanation for the W2 transient in the  $z=0.94$ Warhol arc is a caustic-crossing massive binary system consisting of an RSG and a B-type companion star. The system was detected in four epocchs of JWST NIRCam imaging, and the variation in magnification across the 126-day light curve suggests that the system is being microlensed by a star or other compact object in the $z=0.396$ M0416 cluster. Color differences between epochs require the magnifications of the two stars to vary by differing amounts. The timescale of the microlensing event constrains the projected source-plane size of the system, confirming that W2 is an individual binary system rather than a larger object such as a stellar cluster.  A third microlensing event was subsequently observed at the same position in a single-filter observation acquired $\sim 18$ months later.

Within the binary population-synthesis framework {\tt POSYDON}, $\sim 1$\% of the simulated configurations have the necessary temperatures and luminosities (for maximum magnification $\mu<10000$ ) to reproduce W2's observed SED\null. This translates to $\sim 63$ systems in Warhol's stellar mass of $\sim 10^6$~\Msol. Approximately 46\% of the simulated configurations would be luminous enough at some stage of their evolution to be detected at 5$\sigma$ by the NIRCam observations, so $\sim 3$\% of sufficiently luminous systems go through a phase that could match W2's SED\null.
Some systems can go through this phase twice: first when the more massive star goes through its RSG phase while the secondary remains hot, and a second time if the secondary is in its RSG phase when the primary becomes blue on its way to core collapse.  Other systems match the observations only once.  In some cases, that happens when mass transfer makes the initial secondary the first star to enter its RSG phase.

The probability of observing a system in any particular phase depends on the phase's lifetime and on the probability of achieving the necessary magnification. Taking those into account for systems that potentially match W2, the most likely initial mass of Star~1 is $M1_{\rm init}=23.6^{+5.3}_{-4.3}$~\Msol, and the likely initial mass ratio $q$ is  very close to unity. The initial orbital periods span a wide range, $P1_{\rm init}=5528^{+6372}_{-2145}$ days. Of the possible systems, $\sim 20\%$ experience some amount of mass transfer, but in all cases an RSG still fits within the orbit even after the mass transfer occurs. The typical orbital separations in the weighted sample are consistent with the source-plane size constraints for microlensing.
$\sim37\%$ of the configurations in the weighted sample have a two-phase solution, and the probabilities of observing the system in each phase are approximately equal. If W2 happens to be in this second phase, we are seeing a massive star with a binary companion at $z=0.94$ that is close to exploding as a Type~II supernova. 

Reasonable orbital models can fit W2's observed light curve. For two likely binary configurations we have simulated, 
%from the weighted {\tt POSYDON} (one interacting binary and one non-interacting binary), we are able to reproduce the observed light curve in all filters. Both configurations require 
the center of mass of the binary system must be within $\sim 1$~AU of the microlensing caustic in order to reproduce the light curve, and both configurations favor an eccentric orbit. The orbit solutions have the RSG moving opposite the direction of transverse motion in order to maintain fairly constant magnification, while the blue star orbits in the direction of transverse motion and toward the caustic to produce the observed rapid magnification increase.
However, the transverse velocity is small in comparison to the orbital velocity. The orientation of the tangent to the microcaustic exhibits significant diversity, and the binary orbit matters only when the orbital velocity becomes comparable to the transverse velocity. This may create a selection effect for regions of the microcaustic where the relative motion between the binary and the microlens is close to parallel to the caustic. Pairs of luminous stars are common in the local Universe, and there should be a reasonable probability of observing a binary at small transverse velocities. 
The changing orientations of the microcaustic allow for a large number of angles within the network, and a total of nine events were discovered within the PEARLS/CANUCS observations.

Future monitoring of the Warhol arc can evaluate whether the microlensing event rate at W2's position is in fact exceptionally high. If the rate is very high, then a coincidence of microlensing events instead of a binary system could potentially explain the observations. 

As JWST continues to obtain deep imaging of caustic-crossing lensed galaxies, we expect that many more examples of highly magnified binaries at $z\approx1$ will be discovered. Detecting and analyzing these sources will improve our understanding of massive binaries at high redshifts and their impact on galaxy evolution and of microlens properties in lensing clusters.

\section{Acknowledgments}\label{sec:acknowledgments}
This work is based on observations made with the NASA/ESA/CSA {\it James Webb Space Telescope}. The data were obtained from the Mikulski Archive for Space Telescopes (MAST) at the Space Telescope Science Institute, which is operated by the Association of Universities for Research in Astronomy, Inc., under NASA contract NAS 5-03127 for JWST. These observations are associated with JWST program GO-1176. We thank the CANUCS team for sharing their data. 

P.L.K. acknowledges U.S. National Science Foundation (NSF) AAG program AST-2308051. Support was also from NASA/{HST} grants GO-15936 and GO-16278 from STScI, which is operated by the Association of Universities for Research in Astronomy, Inc., under NASA contract NAS5-26555.  A.V.F. is also grateful for the Christopher R. Redlich Fund and many other donors.  Grant JPL-1659411 provided support for ground-based follow-up observations. E.Z. acknowledges support from the Hellenic Foundation for Research and Innovation (H.F.R.I.) under the ``3rd Call for H.F.R.I. Research Projects to support Post-Doctoral Researchers'' (Project 7933).
R.A.W. acknowledges support from NASA JWST Interdisciplinary Scientist grants NAG5-12460, NNX14AN10G, and 80NSSC18K0200 from GSFC. R.A.W. and S.H.C. acknowledge support from NASA JWST Interdisciplinary Scientist grants NAG5-12460, NNX14AN10G, and 80NSSC18K0200 from GSFC. A.Z. acknowledges support by grant 2020750 from the United States-Israel Binational Science Foundation (BSF) and grant 2109066 from the U.S. NSF, and by Israel Science Foundation Grant 864/23.

\bibliography{bib}{}

@ARTICLE{Kelly_2018,
       author = {{Kelly}, Patrick L. and {Diego}, Jose M. and {Rodney}, Steven and {Kaiser}, Nick and {Broadhurst}, Tom and {Zitrin}, Adi and {Treu}, Tommaso and {P{\'e}rez-Gonz{\'a}lez}, Pablo G. and {Morishita}, Takahiro and {Jauzac}, Mathilde and {Selsing}, Jonatan and {Oguri}, Masamune and {Pueyo}, Laurent and {Ross}, Timothy W. and {Filippenko}, Alexei V. and {Smith}, Nathan and {Hjorth}, Jens and {Cenko}, S. Bradley and {Wang}, Xin and {Howell}, D. Andrew and {Richard}, Johan and {Frye}, Brenda L. and {Jha}, Saurabh W. and {Foley}, Ryan J. and {Norman}, Colin and {Bradac}, Marusa and {Zheng}, Weikang and {Brammer}, Gabriel and {Benito}, Alberto Molino and {Cava}, Antonio and {Christensen}, Lise and {de Mink}, Selma E. and {Graur}, Or and {Grillo}, Claudio and {Kawamata}, Ryota and {Kneib}, Jean-Paul and {Matheson}, Thomas and {McCully}, Curtis and {Nonino}, Mario and {P{\'e}rez-Fournon}, Ismael and {Riess}, Adam G. and {Rosati}, Piero and {Schmidt}, Kasper Borello and {Sharon}, Keren and {Weiner}, Benjamin J.},
        title = "{Extreme magnification of an individual star at redshift 1.5 by a galaxy-cluster lens}",
      journal = {Nature Astronomy},
         year = 2018,
        month = apr,
       volume = {2},
        pages = {334-342},
          doi = {10.1038/s41550-018-0430-3},
archivePrefix = {arXiv},
       eprint = {1706.10279},
 primaryClass = {astro-ph.GA},
       adsurl = {https://ui.adsabs.harvard.edu/abs/2018NatAs...2..334K}
}

@ARTICLE{Rodney_2018,
       author = {{Rodney}, S.~A. and {Balestra}, I. and {Bradac}, M. and {Brammer}, G. and {Broadhurst}, T. and {Caminha}, G.~B. and {Chiriv{\i}}, G. and {Diego}, J.~M. and {Filippenko}, A.~V. and {Foley}, R.~J. and {Graur}, O. and {Grillo}, C. and {Hemmati}, S. and {Hjorth}, J. and {Hoag}, A. and {Jauzac}, M. and {Jha}, S.~W. and {Kawamata}, R. and {Kelly}, P.~L. and {McCully}, C. and {Mobasher}, B. and {Molino}, A. and {Oguri}, M. and {Richard}, J. and {Riess}, A.~G. and {Rosati}, P. and {Schmidt}, K.~B. and {Selsing}, J. and {Sharon}, K. and {Strolger}, L. -G. and {Suyu}, S.~H. and {Treu}, T. and {Weiner}, B.~J. and {Williams}, L.~L.~R. and {Zitrin}, A.},
        title = "{Two peculiar fast transients in a strongly lensed host galaxy}",
      journal = {Nature Astronomy},
         year = 2018,
        month = apr,
       volume = {2},
        pages = {324-333},
          doi = {10.1038/s41550-018-0405-4},
archivePrefix = {arXiv},
       eprint = {1707.02434},
 primaryClass = {astro-ph.GA},
       adsurl = {https://ui.adsabs.harvard.edu/abs/2018NatAs...2..324R}
}

@ARTICLE{Kaurov_2019,
       author = {{Kaurov}, Alexander A. and {Dai}, Liang and {Venumadhav}, Tejaswi and {Miralda-Escud{\'e}}, Jordi and {Frye}, Brenda},
        title = "{Highly Magnified Stars in Lensing Clusters: New Evidence in a Galaxy Lensed by MACS J0416.1-2403}",
      journal = {\apj},
         year = 2019,
        month = jul,
       volume = {880},
       number = {1},
          eid = {58},
        pages = {58},
          doi = {10.3847/1538-4357/ab2888},
archivePrefix = {arXiv},
       eprint = {1902.10090},
 primaryClass = {astro-ph.GA},
       adsurl = {https://ui.adsabs.harvard.edu/abs/2019ApJ...880...58K}
}

@ARTICLE{Chen_2019,
       author = {{Chen}, Wenlei and {Kelly}, Patrick L. and {Diego}, Jose M. and {Oguri}, Masamune and {Williams}, Liliya L.~R. and {Zitrin}, Adi and {Treu}, Tommaso L. and {Smith}, Nathan and {Broadhurst}, Thomas J. and {Kaiser}, Nick and {Foley}, Ryan J. and {Filippenko}, Alexei V. and {Salo}, Laura and {Hjorth}, Jens and {Selsing}, Jonatan},
        title = "{Searching for Highly Magnified Stars at Cosmological Distances: Discovery of a Redshift 0.94 Blue Supergiant in Archival Images of the Galaxy Cluster MACS J0416.1-2403}",
      journal = {\apj},
         year = 2019,
        month = aug,
       volume = {881},
       number = {1},
          eid = {8},
        pages = {8},
          doi = {10.3847/1538-4357/ab297d},
archivePrefix = {arXiv},
       eprint = {1902.05510},
 primaryClass = {astro-ph.GA},
       adsurl = {https://ui.adsabs.harvard.edu/abs/2019ApJ...881....8C}
}

@ARTICLE{Kelly_2022,
       author = {{Kelly}, Patrick L. and {Chen}, Wenlei and {Alfred}, Amruth and {Broadhurst}, Thomas J. and {Diego}, Jose M. and {Emami}, Najmeh and {Filippenko}, Alexei V. and {Keen}, Allison and {Kei Li}, Sung and {Lim}, Jeremy and {Meena}, Ashish K. and {Oguri}, Masamune and {Scarlata}, Claudia and {Treu}, Tommaso and {Williams}, Hayley and {Williams}, Liliya L.~R. and {Zhou}, Rui and {Zitrin}, Adi and {Foley}, Ryan J. and {Jha}, Saurabh W. and {Kaiser}, Nick and {Mehta}, Vihang and {Rieck}, Steven and {Salo}, Laura and {Smith}, Nathan and {Weisz}, Daniel R.},
        title = "{Flashlights: More than A Dozen High-Significance Microlensing Events of Extremely Magnified Stars in Galaxies at Redshifts z=0.7-1.5}",
      journal = {arXiv e-prints},
         year = 2022,
        month = nov,
          eid = {arXiv:2211.02670},
        pages = {arXiv:2211.02670},
          doi = {10.48550/arXiv.2211.02670},
archivePrefix = {arXiv},
       eprint = {2211.02670},
 primaryClass = {astro-ph.CO},
       adsurl = {https://ui.adsabs.harvard.edu/abs/2022arXiv221102670K}
}

@ARTICLE{Yan_2023,
       author = {{Yan}, Haojing and {Ma}, Zhiyuan and {Sun}, Bangzheng and {Wang}, Lifan and {Kelly}, Patrick and {Diego}, Jos{\'e} M. and {Cohen}, Seth H. and {Windhorst}, Rogier A. and {Jansen}, Rolf A. and {Grogin}, Norman A. and {Beacom}, John F. and {Conselice}, Christopher J. and {Driver}, Simon P. and {Frye}, Brenda and {Coe}, Dan and {Marshall}, Madeline A. and {Koekemoer}, Anton and {Willmer}, Christopher N.~A. and {Robotham}, Aaron and {D'Silva}, Jordan C.~J. and {Summers}, Jake and {Nonino}, Mario and {Pirzkal}, Nor and {Ryan}, Russell E. and {Ortiz}, Rafael and {Tompkins}, Scott and {Bhatawdekar}, Rachana A. and {Cheng}, Cheng and {Zitrin}, Adi and {Willner}, S.~P.},
        title = "{JWST's PEARLS: Transients in the MACS J0416.1-2403 Field}",
      journal = {\apjs},
         year = 2023,
        month = dec,
       volume = {269},
       number = {2},
          eid = {43},
        pages = {43},
          doi = {10.3847/1538-4365/ad0298},
archivePrefix = {arXiv},
       eprint = {2307.07579},
 primaryClass = {astro-ph.GA},
       adsurl = {https://ui.adsabs.harvard.edu/abs/2023ApJS..269...43Y}
}

@ARTICLE{Fudamoto_2024,
       author = {{Fudamoto}, Yoshinobu and {Sun}, Fengwu and {Diego}, Jose M. and {Dai}, Liang and {Oguri}, Masamune and {Zitrin}, Adi and {Zackrisson}, Erik and {Jauzac}, Mathilde and {Lagattuta}, David J. and {Egami}, Eiichi and {Iani}, Edoardo and {Windhorst}, Rogier A. and {Abe}, Katsuya T. and {Bauer}, Franz Erik and {Bian}, Fuyan and {Bhatawdekar}, Rachana and {Broadhurst}, Thomas J. and {Cai}, Zheng and {Chen}, Chian-Chou and {Chen}, Wenlei and {Cohen}, Seth H. and {Conselice}, Christopher J. and {Espada}, Daniel and {Foo}, Nicholas and {Frye}, Brenda L. and {Fujimoto}, Seiji and {Furtak}, Lukas J. and {Golubchik}, Miriam and {Hsiao}, Tiger Yu-Yang and {Jolly}, Jean-Baptiste and {Kawai}, Hiroki and {Kelly}, Patrick L. and {Koekemoer}, Anton M. and {Kohno}, Kotaro and {Kokorev}, Vasily and {Li}, Mingyu and {Li}, Zihao and {Lin}, Xiaojing and {Magdis}, Georgios E. and {Meena}, Ashish K. and {Nabizadeh}, Armin and {Richard}, Johan and {Steinhardt}, Charles L. and {Wu}, Yunjing and {Zhu}, Yongda and {Zou}, Siwei},
        title = "{JWST Discovery of $40+$ Microlensed Stars in a Magnified Galaxy, the ``Dragon'' behind Abell 370}",
      journal = {arXiv e-prints},
         year = 2024,
        month = apr,
          eid = {arXiv:2404.08045},
        pages = {arXiv:2404.08045},
          doi = {10.48550/arXiv.2404.08045},
archivePrefix = {arXiv},
       eprint = {2404.08045},
 primaryClass = {astro-ph.GA},
       adsurl = {https://ui.adsabs.harvard.edu/abs/2024arXiv240408045F}
}

@article{Meena_2023,
doi = {10.3847/2041-8213/acb645},
url = {https://dx.doi.org/10.3847/2041-8213/acb645},
year = {2023},
month = {feb},
publisher = {The American Astronomical Society},
volume = {944},
number = {1},
pages = {L6},
author = {Ashish Kumar Meena and Adi Zitrin and Yolanda Jiménez-Teja and Erik Zackrisson and Wenlei Chen and Dan Coe and Jose M. Diego and Paola Dimauro and Lukas J. Furtak and Patrick L. Kelly and Masamune Oguri and Brian Welch and Abdurro’uf  and Felipe Andrade-Santos and Angela Adamo and Rachana Bhatawdekar and Maruša Bradač and Larry D. Bradley and Tom Broadhurst and Christopher J. Conselice and Pratika Dayal and Megan Donahue and Brenda L. Frye and Seiji Fujimoto and Tiger Yu-Yang Hsiao and Vasily Kokorev and Guillaume Mahler and Eros Vanzella and Rogier A. Windhorst},
title = {Two Lensed Star Candidates at z ≃ 4.8 behind the Galaxy Cluster MACS J0647.7+7015},
journal = {The Astrophysical Journal Letters}
}

@ARTICLE{Windhorst_2023,
       author = {{Windhorst}, Rogier A. and {Cohen}, Seth H. and {Jansen}, Rolf A. and {Summers}, Jake and {Tompkins}, Scott and {Conselice}, Christopher J. and {Driver}, Simon P. and {Yan}, Haojing and {Coe}, Dan and {Frye}, Brenda and {Grogin}, Norman and {Koekemoer}, Anton and {Marshall}, Madeline A. and {O'Brien}, Rosalia and {Pirzkal}, Nor and {Robotham}, Aaron and {Ryan}, Russell E. and {Willmer}, Christopher N.~A. and {Carleton}, Timothy and {Diego}, Jose M. and {Keel}, William C. and {Porto}, Paolo and {Redshaw}, Caleb and {Scheller}, Sydney and {Wilkins}, Stephen M. and {Willner}, S.~P. and {Zitrin}, Adi and {Adams}, Nathan J. and {Austin}, Duncan and {Arendt}, Richard G. and {Beacom}, John F. and {Bhatawdekar}, Rachana A. and {Bradley}, Larry D. and {Broadhurst}, Tom and {Cheng}, Cheng and {Civano}, Francesca and {Dai}, Liang and {Dole}, Herv{\'e} and {D'Silva}, Jordan C.~J. and {Duncan}, Kenneth J. and {Fazio}, Giovanni G. and {Ferrami}, Giovanni and {Ferreira}, Leonardo and {Finkelstein}, Steven L. and {Furtak}, Lukas J. and {Gim}, Hansung B. and {Griffiths}, Alex and {Hammel}, Heidi B. and {Harrington}, Kevin C. and {Hathi}, Nimish P. and {Holwerda}, Benne W. and {Honor}, Rachel and {Huang}, Jia-Sheng and {Hyun}, Minhee and {Im}, Myungshin and {Joshi}, Bhavin A. and {Kamieneski}, Patrick S. and {Kelly}, Patrick and {Larson}, Rebecca L. and {Li}, Juno and {Lim}, Jeremy and {Ma}, Zhiyuan and {Maksym}, Peter and {Manzoni}, Giorgio and {Meena}, Ashish Kumar and {Milam}, Stefanie N. and {Nonino}, Mario and {Pascale}, Massimo and {Petric}, Andreea and {Pierel}, Justin D.~R. and {del Carmen Polletta}, Maria and {R{\"o}ttgering}, Huub J.~A. and {Rutkowski}, Michael J. and {Smail}, Ian and {Straughn}, Amber N. and {Strolger}, Louis-Gregory and {Swirbul}, Andi and {Trussler}, James A.~A. and {Wang}, Lifan and {Welch}, Brian and {B. Wyithe}, J. Stuart and {Yun}, Min and {Zackrisson}, Erik and {Zhang}, Jiashuo and {Zhao}, Xiurui},
        title = "{JWST PEARLS. Prime Extragalactic Areas for Reionization and Lensing Science: Project Overview and First Results}",
      journal = {\aj},
         year = 2023,
        month = jan,
       volume = {165},
       number = {1},
          eid = {13},
        pages = {13},
          doi = {10.3847/1538-3881/aca163},
archivePrefix = {arXiv},
       eprint = {2209.04119},
 primaryClass = {astro-ph.CO},
       adsurl = {https://ui.adsabs.harvard.edu/abs/2023AJ....165...13W}
}

@article{Willott_2022,
doi = {10.1088/1538-3873/ac5158},
url = {https://dx.doi.org/10.1088/1538-3873/ac5158},
year = {2022},
month = {feb},
publisher = {The Astronomical Society of the Pacific},
volume = {134},
number = {1032},
pages = {025002},
author = {Chris J. Willott and René Doyon and Loic Albert and Gabriel B. Brammer and William V. Dixon and Koraljka Muzic and Swara Ravindranath and Aleks Scholz and Roberto Abraham and Étienne Artigau and Maruša Bradač and Paul Goudfrooij and John B. Hutchings and Kartheik G. Iyer and Ray Jayawardhana and Stephanie LaMassa and Nicholas Martis and Michael R. Meyer and Takahiro Morishita and Lamiya Mowla and Adam Muzzin and Gaël Noirot and Camilla Pacifici and Neil Rowlands and Ghassan Sarrouh and Marcin Sawicki and Joanna M. Taylor and Kevin Volk and Johannes Zabl},
title = {The Near-infrared Imager and Slitless Spectrograph for the James Webb Space Telescope. II. Wide Field Slitless Spectroscopy},
journal = {Publications of the Astronomical Society of the Pacific}
}

@software{bushouse_2023,
  author       = {Bushouse, Howard and
                  Eisenhamer, Jonathan and
                  Dencheva, Nadia and
                  Davies, James and
                  Greenfield, Perry and
                  Morrison, Jane and
                  Hodge, Phil and
                  Simon, Bernie and
                  Grumm, David and
                  Droettboom, Michael and
                  Slavich, Edward and
                  Sosey, Megan and
                  Pauly, Tyler and
                  Miller, Todd and
                  Jedrzejewski, Robert and
                  Hack, Warren and
                  Davis, David and
                  Crawford, Steven and
                  Law, David and
                  Gordon, Karl and
                  Regan, Michael and
                  Cara, Mihai and
                  MacDonald, Ken and
                  Bradley, Larry and
                  Shanahan, Clare and
                  Jamieson, William and
                  Teodoro, Mairan and
                  Williams, Thomas},
  title        = {JWST Calibration Pipeline},
  month        = jan,
  year         = 2023,
  publisher    = {Zenodo},
  version      = {1.9.4},
  doi          = {10.5281/zenodo.7577320},
  url          = {https://doi.org/10.5281/zenodo.7577320}
}

@ARTICLE{Anderson_2000,
       author = {{Anderson}, Jay and {King}, Ivan R.},
        title = "{Toward High-Precision Astrometry with WFPC2. I. Deriving an Accurate Point-Spread Function}",
      journal = {\pasp},
         year = 2000,
        month = oct,
       volume = {112},
       number = {776},
        pages = {1360-1382},
          doi = {10.1086/316632},
archivePrefix = {arXiv},
       eprint = {astro-ph/0006325},
 primaryClass = {astro-ph},
       adsurl = {https://ui.adsabs.harvard.edu/abs/2000PASP..112.1360A}
}

@software{photutils,
author       = {Larry Bradley and
                Brigitta Sip{\H o}cz and
                Thomas Robitaille and
                Erik Tollerud and
                Z\`e Vin{\'{\i}}cius and
                Christoph Deil and
                Kyle Barbary and
                Tom J Wilson and
                Ivo Busko and
                Axel Donath and
                Hans Moritz G{\"u}nther and
                Mihai Cara and
                P. L. Lim and
                Sebastian Me{\ss}linger and
                Simon Conseil and
                Azalee Bostroem and
                Michael Droettboom and
                E. M. Bray and
                Lars Andersen Bratholm and
                Geert Barentsen and
                Matt Craig and
                Shivangee Rathi and
                Sergio Pascual and
                Gabriel Perren and
                Iskren Y. Georgiev and
                Miguel de Val-Borro and
                Wolfgang Kerzendorf and
                Yoonsoo P. Bach and
                Bruno Quint and
                Harrison Souchereau},
title        = {astropy/photutils: 1.8.0},
month        = may,
year         = 2023,
publisher    = {Zenodo},
version      = {1.8.0},
doi          = {10.5281/zenodo.7946442},
url          = {https://doi.org/10.5281/zenodo.7946442}
}

@article{Sana_2013,
	author = {{Sana, H.} and {de Koter, A.} and {de Mink, S. E.} and {Dunstall, P. R.} and {Evans, C. J.} and {Hénault-Brunet, V.} and {Maíz Apellániz, J.} and {Ramírez-Agudelo, O. H.} and {Taylor, W. D.} and {Walborn, N. R.} and {Clark, J. S.} and {Crowther, P. A.} and {Herrero, A.} and {Gieles, M.} and {Langer, N.} and {Lennon, D. J.} and {Vink, J. S.}},
	title = {The VLT-FLAMES Tarantula Survey ⋆⋆⋆⋆⋆⋆ - VIII. Multiplicity properties of the O-type star
          population},
	DOI= "10.1051/0004-6361/201219621",
	url= "https://doi.org/10.1051/0004-6361/201219621",
	journal = {A\&A},
	year = 2013,
	volume = 550,
	pages = "A107",
	month = "",
}

@ARTICLE{Guo_2022,
  author = {{Guo}, Yanjun and {Liu}, Chao and {Wang}, Luqian and {Wang}, Jinliang and {Zhang}, Bo and {Ji}, Kaifan and {Han}, ZhanWen and {Chen}, XueFei},
  title = "{The statistical properties of early-type stars from LAMOST DR8}",
  journal = {\aap},
  year = 2022,
  month = nov,
  volume = {667},
  eid = {A44},
  pages = {A44},
  doi = {10.1051/0004-6361/202244300},
  archivePrefix = {arXiv},
  eprint = {2209.09272},
  primaryClass = {astro-ph.SR}
}

@ARTICLE{Moe_2017,
       author = {{Moe}, Maxwell and {Di Stefano}, Rosanne},
        title = "{Mind Your Ps and Qs: The Interrelation between Period (P) and Mass-ratio (Q) Distributions of Binary Stars}",
      journal = {\apjs},
         year = 2017,
        month = jun,
       volume = {230},
       number = {2},
          eid = {15},
        pages = {15},
          doi = {10.3847/1538-4365/aa6fb6},
archivePrefix = {arXiv},
       eprint = {1606.05347},
 primaryClass = {astro-ph.SR},
       adsurl = {https://ui.adsabs.harvard.edu/abs/2017ApJS..230...15M}
}

@ARTICLE{Frost_2025,
       author = {{Frost}, A.~J. and {Sana}, H. and {Le Bouquin}, J-B and {Perets}, H.~B. and {Bodensteiner}, J. and {Igoshev}, A.~P. and {Banyard}, G. and {Mahy}, L. and {M{\'e}rand}, A. and {Ram{\'\i}rez-Agudelo}, O.~H.},
        title = "{An interferometric study of B star multiplicity}",
      journal = {arXiv e-prints},
         year = 2025,
        month = may,
          eid = {arXiv:2505.02300},
        pages = {arXiv:2505.02300},
          doi = {10.48550/arXiv.2505.02300},
archivePrefix = {arXiv},
       eprint = {2505.02300},
 primaryClass = {astro-ph.SR},
       adsurl = {https://ui.adsabs.harvard.edu/abs/2025arXiv250502300F}
}

@misc{Dai_2025,
      title={The Binary Fraction of Red Supergiants in the Magellanic Clouds}, 
      author={Min Dai and Shu Wang and Biwei Jiang},
      year={2025},
      eprint={2504.03357},
      archivePrefix={arXiv},
      primaryClass={astro-ph.SR},
      url={https://arxiv.org/abs/2504.03357}, 
}

@ARTICLE{Patrick_2022,
       author = {{Patrick}, L.~R. and {Thilker}, D. and {Lennon}, D.~J. and {Bianchi}, L. and {Schootemeijer}, A. and {Dorda}, R. and {Langer}, N. and {Negueruela}, I.},
        title = "{Red supergiant stars in binary systems. I. Identification and characterization in the small magellanic cloud from the UVIT ultraviolet imaging survey}",
      journal = {\mnras},
         year = 2022,
        month = jul,
       volume = {513},
       number = {4},
        pages = {5847-5860},
          doi = {10.1093/mnras/stac1139},
archivePrefix = {arXiv},
       eprint = {2204.11866},
 primaryClass = {astro-ph.SR},
       adsurl = {https://ui.adsabs.harvard.edu/abs/2022MNRAS.513.5847P}
}

@article{PatrickNeguerela2024,
doi={10.25518/0037-9565.12310},
url={https://popups.uliege.be/0037-9565/index.php?id=12310},
year={2024},
month={July},
publisher={Bulletin de la Société Royale des Sciences de Liège},
volume = {93},
number = {3},
pages={173-192},
journal={Bull. Soc. R. Sci},
title={Hot and Cool: Characterising the Companions of Red Supergiant Stars in Binary Systems},
author={ Patrick, Lee R. and Negueruela, Ignacio}}

@article{Neugent_2020,
doi = {10.3847/1538-4357/ababaa},
url = {https://dx.doi.org/10.3847/1538-4357/ababaa},
year = {2020},
month = {sep},
publisher = {The American Astronomical Society},
volume = {900},
number = {2},
pages = {118},
author = {Neugent, Kathryn F. and Levesque, Emily M. and Massey, Philip and Morrell, Nidia I. and Drout, Maria R.},
title = {The Red Supergiant Binary Fraction of the Large Magellanic Cloud},
journal = {The Astrophysical Journal}
}

@article{Sana_2012,
   title={Binary Interaction Dominates the Evolution of Massive Stars},
   volume={337},
   ISSN={1095-9203},
   url={http://dx.doi.org/10.1126/science.1223344},
   DOI={10.1126/science.1223344},
   number={6093},
   journal={Science},
   publisher={American Association for the Advancement of Science (AAAS)},
   author={Sana, H. and de Mink, S. E. and de Koter, A. and Langer, N. and Evans, C. J. and Gieles, M. and Gosset, E. and Izzard, R. G. and Le Bouquin, J.-B. and Schneider, F. R. N.},
   year={2012},
   month=jul, pages={444–446} }

@ARTICLE{Humphreys_1979,
       author = {{Humphreys}, R.~M. and {Davidson}, K.},
        title = "{Studies of luminous stars in nearby galaxies. III. Comments on the evolution of the most massive stars in the Milky Way and the Large Magellanic Cloud.}",
      journal = {\apj},
         year = 1979,
        month = sep,
       volume = {232},
        pages = {409-420},
          doi = {10.1086/157301},
       adsurl = {https://ui.adsabs.harvard.edu/abs/1979ApJ...232..409H}
}

@article{Davies_2013,
   title={THE TEMPERATURES OF RED SUPERGIANTS},
   volume={767},
   ISSN={1538-4357},
   url={http://dx.doi.org/10.1088/0004-637X/767/1/3},
   DOI={10.1088/0004-637x/767/1/3},
   number={1},
   journal={The Astrophysical Journal},
   publisher={American Astronomical Society},
   author={Davies, Ben and Kudritzki, Rolf-Peter and Plez, Bertrand and Trager, Scott and Lançon, Ariane and Gazak, Zach and Bergemann, Maria and Evans, Chris and Chiavassa, Andrea},
   year={2013},
   month=mar, pages={3} }

@ARTICLE{Levesque_2010,
       author = {{Levesque}, Emily M.},
        title = "{The physical properties of red supergiants}",
      journal = {\nar},
         year = 2010,
        month = jan,
       volume = {54},
       number = {1-2},
        pages = {1-12},
          doi = {10.1016/j.newar.2009.10.002},
archivePrefix = {arXiv},
       eprint = {0902.2789},
 primaryClass = {astro-ph.SR},
       adsurl = {https://ui.adsabs.harvard.edu/abs/2010NewAR..54....1L}
}

@article{Doughty_2021,
   title={The effects of binary stars on galaxies and metal-enriched gas during reionization},
   volume={505},
   ISSN={1365-2966},
   url={http://dx.doi.org/10.1093/mnras/stab1448},
   DOI={10.1093/mnras/stab1448},
   number={2},
   journal={Monthly Notices of the Royal Astronomical Society},
   publisher={Oxford University Press (OUP)},
   author={Doughty, Caitlin and Finlator, Kristian},
   year={2021},
   month=may, pages={2207–2223} }

@article{deSa_2024,
   title={Compact object populations over cosmic time II. Compact object merger rates and masses over redshift from varying initial conditions},
   volume={535},
   ISSN={1365-2966},
   url={http://dx.doi.org/10.1093/mnras/stae2281},
   DOI={10.1093/mnras/stae2281},
   number={3},
   journal={Monthly Notices of the Royal Astronomical Society},
   publisher={Oxford University Press (OUP)},
   author={de Sá, Lucas M and Rocha, Lívia S and Bernardo, Antônio and Bachega, Riis R A and Horvath, Jorge E},
   year={2024},
   month=nov, pages={2041–2067} }

@article{Eldridge_2018, title={Supernova lightCURVE POPulation Synthesis I: Including interacting binaries is key to understanding the diversity of type II supernova lightcurves}, volume={35}, DOI={10.1017/pasa.2018.47}, journal={Publications of the Astronomical Society of Australia}, author={Eldridge, J. J. and Xiao, L. and Stanway, E. R. and Rodrigues, N. and Guo, N.-Y.}, year={2018}, pages={e049}}

@ARTICLE{Fragos_2023,
       author = {{Fragos}, Tassos and {Andrews}, Jeff J. and {Bavera}, Simone S. and {Berry}, Christopher P.~L. and {Coughlin}, Scott and {Dotter}, Aaron and {Giri}, Prabin and {Kalogera}, Vicky and {Katsaggelos}, Aggelos and {Kovlakas}, Konstantinos and {Lalvani}, Shamal and {Misra}, Devina and {Srivastava}, Philipp M. and {Qin}, Ying and {Rocha}, Kyle A. and {Rom{\'a}n-Garza}, Jaime and {Serra}, Juan Gabriel and {Stahle}, Petter and {Sun}, Meng and {Teng}, Xu and {Trajcevski}, Goce and {Tran}, Nam Hai and {Xing}, Zepei and {Zapartas}, Emmanouil and {Zevin}, Michael},
        title = "{POSYDON: A General-purpose Population Synthesis Code with Detailed Binary-evolution Simulations}",
      journal = {\apjs},
         year = 2023,
        month = feb,
       volume = {264},
       number = {2},
          eid = {45},
        pages = {45},
          doi = {10.3847/1538-4365/ac90c1},
archivePrefix = {arXiv},
       eprint = {2202.05892},
 primaryClass = {astro-ph.SR},
       adsurl = {https://ui.adsabs.harvard.edu/abs/2023ApJS..264...45F}
}

@article{KassRaftery_1995,
year={1995},
month={June},
journal={Journal of the American Statistical Association},
volume={90},
number={430},
pages={773-795},
author={Kass, Robert E. and Raftery, Adrian E.},
title={Bayes Factors}
}

@ARTICLE{Pollmann_2020,
       author = {{Pollmann}, E. and {Bennett}, P.},
        title = "{Spectroscopic Monitoring of the 2017-2019 Eclipse of VV Cephei}",
      journal = {\jaavso},
         year = 2020,
        month = dec,
       volume = {48},
       number = {2},
        pages = {118},
       adsurl = {https://ui.adsabs.harvard.edu/abs/2020JAVSO..48..118P}
}

@ARTICLE{Pollmann_2023,
       author = {{Pollmann}, Ernst and {Bennett}, Philip},
        title = "{Spectroscopic Monitoring of the 2017{\textendash}2019 Eclipse of VV Cephei}",
      journal = {BAV Magazine Spectroscopy},
         year = 2023,
        month = jan,
       volume = {13},
        pages = {12-23},
       adsurl = {https://ui.adsabs.harvard.edu/abs/2023BAVMS..13...12P}
}

@ARTICLE{Wright_1977,
       author = {{Wright}, K.~O.},
        title = "{The System of VV Cephei Derived from an Analysis of the H{\ensuremath{\alpha}} Line}",
      journal = {\jrasc},
         year = 1977,
        month = apr,
       volume = {71},
        pages = {152},
       adsurl = {https://ui.adsabs.harvard.edu/abs/1977JRASC..71..152W}
}

@ARTICLE{Cowley_1965,
       author = {{Cowley}, Anne Pyne},
        title = "{A Spectroscopic Study of the Peculiar Binary Boss 1985.}",
      journal = {\apj},
         year = 1965,
        month = jul,
       volume = {142},
        pages = {299},
          doi = {10.1086/148284},
       adsurl = {https://ui.adsabs.harvard.edu/abs/1965ApJ...142..299C}
}

@article{Jaschek_1963,
doi = {10.1086/128017},
url = {https://doi.org/10.1086/128017},
year = {1963},
month = {dec},
publisher = {The Astronomical Society of the Pacific},
volume = {75},
number = {447},
pages = {509},
author = {Jaschek, Carlos and Jaschek, Mercedes},
title = {THE SPECTRUM OF HR 2902 IN 1961},
journal = {Publications of the Astronomical Society of the Pacific}
}

@ARTICLE{Rossi_1992
,
       author = {{Rossi}, C. and {Altamore}, A. and {Baratta}, G.~B. and {Friedjung}, M. and {Viotti}, R.},
        title = "{The spectrum of the VV Cephei star KQ Puppis (Boss 1985). III. A possible model.}",
      journal = {\aap},
         year = 1992,
        month = mar,
       volume = {256},
        pages = {133-140},
       adsurl = {https://ui.adsabs.harvard.edu/abs/1992A&A...256..133R}
}

@book{Thomas_2015,
author = {Ake, Thomas and Griffin, Elizabeth},
year = {2015},
month = {01},
pages = {},
title = {Giants of Eclipse: The $\zeta$ Aurigae Stars and Other Binary Systems},
isbn = {978-3-319-09197-6},
doi = {10.1007/978-3-319-09198-3}
}

@article{Paxton_2015,
doi = {10.1088/0067-0049/220/1/15},
url = {https://dx.doi.org/10.1088/0067-0049/220/1/15},
year = {2015},
month = {sep},
publisher = {The American Astronomical Society},
volume = {220},
number = {1},
pages = {15},
author = {Paxton, Bill and Marchant, Pablo and Schwab, Josiah and Bauer, Evan B. and Bildsten, Lars and Cantiello, Matteo and Dessart, Luc and Farmer, R. and Hu, H. and Langer, N. and Townsend, R. H. D. and Townsley, Dean M. and Timmes, F. X.},
title = {MODULES FOR EXPERIMENTS IN STELLAR ASTROPHYSICS (MESA): BINARIES, PULSATIONS, AND EXPLOSIONS},
journal = {The Astrophysical Journal Supplement Series}
}

@ARTICLE{Lejeune_1998,
       author = {{Lejeune}, T. and {Cuisinier}, F. and {Buser}, R.},
        title = "{VizieR Online Data Catalog: A standard stellar library. II. (Lejeune+ 1998)}",
      journal = {VizieR Online Data Catalog},
         year = 1998,
        month = jun,
          eid = {J/A+AS/130/65},
        pages = {J/A+AS/130/65},
       adsurl = {https://ui.adsabs.harvard.edu/abs/1998yCat..41300065L}
}

@article{schwarz1978estimating,
  title={Estimating the dimension of a model},
  author={Schwarz, Gideon},
  journal={The annals of statistics},
  pages={461--464},
  year={1978},
  publisher={JSTOR}
}

@ARTICLE{Miralda_1991,
       author = {{Miralda-Escude}, Jordi},
        title = "{The Magnification of Stars Crossing a Caustic. I. Lenses with Smooth Potentials}",
      journal = {\apj},
         year = 1991,
        month = sep,
       volume = {379},
        pages = {94},
          doi = {10.1086/170486},
       adsurl = {https://ui.adsabs.harvard.edu/abs/1991ApJ...379...94M}
}

@ARTICLE{Kroupa_2001,
       author = {{Kroupa}, Pavel},
        title = "{On the variation of the initial mass function}",
      journal = {\mnras},
         year = 2001,
        month = apr,
       volume = {322},
       number = {2},
        pages = {231-246},
          doi = {10.1046/j.1365-8711.2001.04022.x},
archivePrefix = {arXiv},
       eprint = {astro-ph/0009005},
 primaryClass = {astro-ph},
       adsurl = {https://ui.adsabs.harvard.edu/abs/2001MNRAS.322..231K}
}

@ARTICLE{Palencia_2024,
       author = {{Palencia}, J.~M. and {Diego}, J.~M. and {Kavanagh}, B.~J. and {Mart{\'\i}nez-Arrizabalaga}, J.},
        title = "{Statistics of magnification for extremely lensed high redshift stars}",
      journal = {\aap},
         year = 2024,
        month = jul,
       volume = {687},
          eid = {A81},
        pages = {A81},
          doi = {10.1051/0004-6361/202347492},
archivePrefix = {arXiv},
       eprint = {2307.09505},
 primaryClass = {astro-ph.CO},
       adsurl = {https://ui.adsabs.harvard.edu/abs/2024A&A...687A..81P}
}

@article{kawamataoguriishigaki16,
	adsurl = {http://adsabs.harvard.edu/abs/2016ApJ...819..114K},
	archiveprefix = {arXiv},
	author = {{Kawamata}, R. and {Oguri}, M. and {Ishigaki}, M. and {Shimasaku}, K. and {Ouchi}, M.},
	doi = {10.3847/0004-637X/819/2/114},
	eid = {114},
	eprint = {1510.06400},
	journal = {\apj},
	month = mar,
	pages = {114},
	title = {{Precise Strong Lensing Mass Modeling of Four Hubble Frontier Field Clusters and a Sample of Magnified High-redshift Galaxies}},
	volume = 819,
	year = 2016}

@article{kawamataishigakishimasaku18,
	adsurl = {https://ui.adsabs.harvard.edu/abs/2018ApJ...855....4K},
	archiveprefix = {arXiv},
	author = {{Kawamata}, Ryota and {Ishigaki}, Masafumi and {Shimasaku}, Kazuhiro and {Oguri}, Masamune and {Ouchi}, Masami and {Tanigawa}, Shingo},
	doi = {10.3847/1538-4357/aaa6cf},
	eid = {4},
	eprint = {1710.07301},
	journal = {\apj},
	month = mar,
	number = {1},
	pages = {4},
	primaryclass = {astro-ph.GA},
	title = {{Size-Luminosity Relations and UV Luminosity Functions at z = 6-9 Simultaneously Derived from the Complete Hubble Frontier Fields Data}},
	volume = {855},
	year = 2018}

@article{oguri10,
	adsurl = {http://adsabs.harvard.edu/abs/2010PASJ...62.1017O},
	archiveprefix = {arXiv},
	author = {{Oguri}, M.},
	doi = {10.1093/pasj/62.4.1017},
	eprint = {1005.3103},
	journal = {\pasj},
	month = aug,
	pages = {1017-1024},
	title = {{The Mass Distribution of SDSS J1004+4112 Revisited}},
	volume = 62,
	year = 2010}

@ARTICLE{Kayser_1986,
       author = {{Kayser}, R. and {Refsdal}, S. and {Stabell}, R.},
        title = "{Astrophysical applications of gravitational micro-lensing.}",
      journal = {\aap},
         year = 1986,
        month = sep,
       volume = {166},
        pages = {36-52},
       adsurl = {https://ui.adsabs.harvard.edu/abs/1986A&A...166...36K}
}

@ARTICLE{Kogut_1993,
       author = {{Kogut}, A. and {Lineweaver}, C. and {Smoot}, G.~F. and {Bennett}, C.~L. and {Banday}, A. and {Boggess}, N.~W. and {Cheng}, E.~S. and {de Amici}, G. and {Fixsen}, D.~J. and {Hinshaw}, G. and {Jackson}, P.~D. and {Janssen}, M. and {Keegstra}, P. and {Loewenstein}, K. and {Lubin}, P. and {Mather}, J.~C. and {Tenorio}, L. and {Weiss}, R. and {Wilkinson}, D.~T. and {Wright}, E.~L.},
        title = "{Dipole Anisotropy in the COBE Differential Microwave Radiometers First-Year Sky Maps}",
      journal = {\apj},
         year = 1993,
        month = dec,
       volume = {419},
        pages = {1},
          doi = {10.1086/173453},
archivePrefix = {arXiv},
       eprint = {astro-ph/9312056},
 primaryClass = {astro-ph},
       adsurl = {https://ui.adsabs.harvard.edu/abs/1993ApJ...419....1K}
}

@ARTICLE{Vovk_2016,
       author = {{Vovk}, Ie. and {Neronov}, A.},
        title = "{Microlensing constraints on the size of the gamma-ray emission region in blazar B0218+357}",
      journal = {\aap},
         year = 2016,
        month = feb,
       volume = {586},
          eid = {A150},
        pages = {A150},
          doi = {10.1051/0004-6361/201526918},
archivePrefix = {arXiv},
       eprint = {1507.01092},
 primaryClass = {astro-ph.HE},
       adsurl = {https://ui.adsabs.harvard.edu/abs/2016A&A...586A.150V}
}

@ARTICLE{Ryon_2015,
       author = {{Ryon}, J.~E. and {Bastian}, N. and {Adamo}, A. and {Konstantopoulos}, I.~S. and {Gallagher}, J.~S. and {Larsen}, S. and {Hollyhead}, K. and {Silva-Villa}, E. and {Smith}, L.~J.},
        title = "{Sizes and shapes of young star cluster light profiles in M83}",
      journal = {\mnras},
         year = 2015,
        month = sep,
       volume = {452},
       number = {1},
        pages = {525-539},
          doi = {10.1093/mnras/stv1282},
archivePrefix = {arXiv},
       eprint = {1506.02042},
 primaryClass = {astro-ph.GA},
       adsurl = {https://ui.adsabs.harvard.edu/abs/2015MNRAS.452..525R}
}

@ARTICLE{Zheng_2025,
       author = {{Zheng}, Wenwen and {Fu}, Xiaoting and {Chen}, Yang and {Chen}, Xuefei and {Guo}, Yanjun and {Chen}, Xuechun and {Shan}, Huanyuan and {Li}, Guoliang},
        title = "{The Magnified Waltz: Simulating Light Curves of Binary Stars Passing through Micro-Caustics in Strong Lensing Galaxy Clusters}",
      journal = {arXiv e-prints},
         year = 2025,
        month = mar,
          eid = {arXiv:2503.04899},
        pages = {arXiv:2503.04899},
          doi = {10.48550/arXiv.2503.04899},
archivePrefix = {arXiv},
       eprint = {2503.04899},
 primaryClass = {astro-ph.GA},
       adsurl = {https://ui.adsabs.harvard.edu/abs/2025arXiv250304899Z}
}

@ARTICLE{Palencia_2025,
       author = {{Palencia}, J.~M. and {Diego}, J.~M. and {Dai}, L. and {Pascale}, M. and {Windhorst}, R. and {Koekemoer}, A.~M. and {Li}, Sung Kei and {Kavanagh}, B.~J. and {Sun}, Fengwu and {Alfred}, Amruth and {Meena}, Ashish K. and {Broadhurst}, Thomas J. and {Kelly}, Patrick L. and {Perera}, Derek and {Williams}, Hayley and {Zitrin}, Adi},
        title = "{Microlensing at Cosmological Distances: Event Rate Predictions in the Warhol Arc of MACS 0416}",
      journal = {arXiv e-prints},
         year = 2025,
        month = apr,
          eid = {arXiv:2504.07039},
        pages = {arXiv:2504.07039},
          doi = {10.48550/arXiv.2504.07039},
archivePrefix = {arXiv},
       eprint = {2504.07039},
 primaryClass = {astro-ph.CO},
       adsurl = {https://ui.adsabs.harvard.edu/abs/2025arXiv250407039P}
}

@ARTICLE{Windhorst_2018,
       author = {{Windhorst}, Rogier A. and {Timmes}, F.~X. and {Wyithe}, J. Stuart B. and {Alpaslan}, Mehmet and {Andrews}, Stephen K. and {Coe}, Daniel and {Diego}, Jose M. and {Dijkstra}, Mark and {Driver}, Simon P. and {Kelly}, Patrick L. and {Kim}, Duho},
        title = "{On the Observability of Individual Population III Stars and Their Stellar-mass Black Hole Accretion Disks through Cluster Caustic Transits}",
      journal = {\apjs},
         year = 2018,
        month = feb,
       volume = {234},
       number = {2},
          eid = {41},
        pages = {41},
          doi = {10.3847/1538-4365/aaa760},
archivePrefix = {arXiv},
       eprint = {1801.03584},
 primaryClass = {astro-ph.GA},
       adsurl = {https://ui.adsabs.harvard.edu/abs/2018ApJS..234...41W}
}

@ARTICLE{Banyard+2022,
       author = {{Banyard}, G. and {Sana}, H. and {Mahy}, L. and {Bodensteiner}, J. and {Villase{\~n}or}, J.~I. and {Evans}, C.~J.},
        title = "{The observed multiplicity properties of B-type stars in the Galactic young open cluster NGC 6231}",
      journal = {\aap},
         year = 2022,
        month = feb,
       volume = {658},
          eid = {A69},
        pages = {A69},
          doi = {10.1051/0004-6361/202141037},
archivePrefix = {arXiv},
       eprint = {2108.07814},
 primaryClass = {astro-ph.SR},
       adsurl = {https://ui.adsabs.harvard.edu/abs/2022A&A...658A..69B}
}

@ARTICLE{Gotberg_2020,
       author = {{G{\"o}tberg}, Y. and {de Mink}, S.~E. and {McQuinn}, M. and {Zapartas}, E. and {Groh}, J.~H. and {Norman}, C.},
        title = "{Contribution from stars stripped in binaries to cosmic reionization of hydrogen and helium}",
      journal = {\aap},
         year = 2020,
        month = feb,
       volume = {634},
          eid = {A134},
        pages = {A134},
          doi = {10.1051/0004-6361/201936669},
archivePrefix = {arXiv},
       eprint = {1911.00543},
 primaryClass = {astro-ph.GA},
       adsurl = {https://ui.adsabs.harvard.edu/abs/2020A&A...634A.134G}
}

@ARTICLE{Paxton2011,
  author = {{Paxton}, B. and {Bildsten}, L. and {Dotter}, A. and {Herwig}, F. and {Lesaffre}, P. and {Timmes}, F.},
  title = {{Modules for Experiments in Stellar Astrophysics (MESA)}},
  journal = {\apjs},
  archivePrefix = {arXiv},
  eprint = {1009.1622},
  primaryClass = {astro-ph.SR},
  year = {2011},
  month = {jan},
  volume = {192},
  eid = {3},
  pages = {3},
  doi = {10.1088/0067-0049/192/1/3},
  adsurl = {https://ui.adsabs.harvard.edu/abs/2011ApJS..192....3P},
}

@ARTICLE{Paxton2013,
  author = {{Paxton}, B. and {Cantiello}, M. and {Arras}, P. and {Bildsten}, L. and {Brown}, E.~F. and {Dotter}, A. and {Mankovich}, C. and {Montgomery}, M.~H. and {Stello}, D. and {Timmes}, F.~X. and {Townsend}, R.},
  title = {{Modules for Experiments in Stellar Astrophysics (MESA): Planets, Oscillations, Rotation, and Massive Stars}},
  journal = {\apjs},
  archivePrefix = {arXiv},
  eprint = {1301.0319},
  primaryClass = {astro-ph.SR},
  year = {2013},
  month = {sep},
  volume = {208},
  eid = {4},
  pages = {4},
  doi = {10.1088/0067-0049/208/1/4},
  adsurl = {https://ui.adsabs.harvard.edu/abs/2013ApJS..208....4P},
}

@ARTICLE{Paxton2018,
  author = {{Paxton}, B. and {Schwab}, J. and {Bauer}, E.~B. and {Bildsten}, L. and {Blinnikov}, S. and {Duffell}, P. and {Farmer}, R. and {Goldberg}, J.~A. and {Marchant}, P. and {Sorokina}, E. and {Thoul}, A. and {Townsend}, R.~H.~D. and {Timmes}, F.~X.},
  title = {{Modules for Experiments in Stellar Astrophysics (MESA): Convective Boundaries, Element Diffusion, and Massive Star Explosions}},
  journal = {\apjs},
  archivePrefix = {arXiv},
  eprint = {1710.08424},
  primaryClass = {astro-ph.SR},
  year = {2018},
  month = {feb},
  volume = {234},
  eid = {34},
  pages = {34},
  doi = {10.3847/1538-4365/aaa5a8},
  adsurl = {https://ui.adsabs.harvard.edu/abs/2018ApJS..234...34P},
}

@ARTICLE{Paxton2019,
       author = {{Paxton}, Bill and {Smolec}, R. and {Schwab}, Josiah and {Gautschy}, A. and
         {Bildsten}, Lars and {Cantiello}, Matteo and {Dotter}, Aaron and
         {Farmer}, R. and {Goldberg}, Jared A. and {Jermyn}, Adam S. and
         {Kanbur}, S.~M. and {Marchant}, Pablo and {Thoul}, Anne and
         {Townsend}, Richard H.~D. and {Wolf}, William M. and {Zhang}, Michael and
         {Timmes}, F.~X.},
        title = "{Modules for Experiments in Stellar Astrophysics (MESA): Pulsating Variable Stars, Rotation, Convective Boundaries, and Energy Conservation}",
      journal = {\apjs},
         year = "2019",
        month = "Jul",
       volume = {243},
       number = {1},
          eid = {10},
        pages = {10},
          doi = {10.3847/1538-4365/ab2241},
archivePrefix = {arXiv},
       eprint = {1903.01426},
 primaryClass = {astro-ph.SR},
       adsurl = {https://ui.adsabs.harvard.edu/abs/2019ApJS..243...10P}
}

@ARTICLE{Niejsel_2019,
       author = {{Neijssel}, Coenraad J. and {Vigna-G{\'o}mez}, Alejandro and {Stevenson}, Simon and {Barrett}, Jim W. and {Gaebel}, Sebastian M. and {Broekgaarden}, Floor S. and {de Mink}, Selma E. and {Sz{\'e}csi}, Dorottya and {Vinciguerra}, Serena and {Mandel}, Ilya},
        title = "{The effect of the metallicity-specific star formation history on double compact object mergers}",
      journal = {\mnras},
         year = 2019,
        month = dec,
       volume = {490},
       number = {3},
        pages = {3740-3759},
          doi = {10.1093/mnras/stz2840},
archivePrefix = {arXiv},
       eprint = {1906.08136},
 primaryClass = {astro-ph.SR},
       adsurl = {https://ui.adsabs.harvard.edu/abs/2019MNRAS.490.3740N}
}

@ARTICLE{Bavera_2021,
       author = {{Bavera}, Simone S. and {Fragos}, Tassos and {Zevin}, Michael and {Berry}, Christopher P.~L. and {Marchant}, Pablo and {Andrews}, Jeff J. and {Coughlin}, Scott and {Dotter}, Aaron and {Kovlakas}, Konstantinos and {Misra}, Devina and {Serra-Perez}, Juan G. and {Qin}, Ying and {Rocha}, Kyle A. and {Rom{\'a}n-Garza}, Jaime and {Tran}, Nam H. and {Zapartas}, Emmanouil},
        title = "{The impact of mass-transfer physics on the observable properties of field binary black hole populations}",
      journal = {\aap},
         year = 2021,
        month = mar,
       volume = {647},
          eid = {A153},
        pages = {A153},
          doi = {10.1051/0004-6361/202039804},
archivePrefix = {arXiv},
       eprint = {2010.16333},
 primaryClass = {astro-ph.HE},
       adsurl = {https://ui.adsabs.harvard.edu/abs/2021A&A...647A.153B}
}

@ARTICLE{Klencki_2022,
       author = {{Klencki}, Jakub and {Istrate}, Alina and {Nelemans}, Gijs and {Pols}, Onno},
        title = "{Partial-envelope stripping and nuclear-timescale mass transfer from evolved supergiants at low metallicity}",
      journal = {\aap},
         year = 2022,
        month = jun,
       volume = {662},
          eid = {A56},
        pages = {A56},
          doi = {10.1051/0004-6361/202142701},
archivePrefix = {arXiv},
       eprint = {2111.10271},
 primaryClass = {astro-ph.SR},
       adsurl = {https://ui.adsabs.harvard.edu/abs/2022A&A...662A..56K}
}

@article{Reid_2002,
doi = {10.1086/338947},
url = {https://dx.doi.org/10.1086/338947},
year = {2002},
month = {apr},
publisher = {},
volume = {568},
number = {2},
pages = {931},
author = {Reid, M. J. and Peek, J. E. G.},
title = {How Mira Variables Change Visual Light by a Thousandfold},
journal = {The Astrophysical Journal}
}

@ARTICLE{Weis_2020,
       author = {{Weis}, Kerstin and {Bomans}, Dominik J.},
        title = "{Luminous Blue Variables}",
      journal = {Galaxies},
         year = 2020,
        month = feb,
       volume = {8},
       number = {1},
          eid = {20},
        pages = {20},
          doi = {10.3390/galaxies8010020},
archivePrefix = {arXiv},
       eprint = {2009.03144},
 primaryClass = {astro-ph.SR},
       adsurl = {https://ui.adsabs.harvard.edu/abs/2020Galax...8...20W}
}

@INPROCEEDINGS{Sterken_2003,
       author = {{Sterken}, C.},
        title = "{Cycles and cyclicities in Luminous Blue Variables: the S Dor phenomenon}",
    booktitle = {Interplay of Periodic, Cyclic and Stochastic Variability in Selected Areas of the H-R Diagram},
         year = 2003,
       editor = {{Sterken}, C.},
       series = {Astronomical Society of the Pacific Conference Series},
       volume = {292},
        month = mar,
        pages = {437},
       adsurl = {https://ui.adsabs.harvard.edu/abs/2003ASPC..292..437S}
}

@ARTICLE{Maravelias_2023,
       author = {{Maravelias}, Grigoris and {de Wit}, Stephan and {Bonanos}, Alceste Z. and {Tramper}, Frank and {Munoz-Sanchez}, Gonzalo and {Christodoulou}, Evangelia},
        title = "{Discovering New B[e] Supergiants and Candidate Luminous Blue Variables in Nearby Galaxies}",
      journal = {Galaxies},
         year = 2023,
        month = jun,
       volume = {11},
       number = {3},
          eid = {79},
        pages = {79},
          doi = {10.3390/galaxies11030079},
archivePrefix = {arXiv},
       eprint = {2307.03320},
 primaryClass = {astro-ph.GA},
       adsurl = {https://ui.adsabs.harvard.edu/abs/2023Galax..11...79M}
}

@article{Aret_2012,
    author = {Aret, A. and Kraus, M. and Muratore, M. F. and Borges Fernandes, M.},
    title = {A new observational tracer for high-density disc-like structures around B[e] supergiants*},
    journal = {Monthly Notices of the Royal Astronomical Society},
    volume = {423},
    number = {1},
    pages = {284-293},
    year = {2012},
    month = {05},
    issn = {0035-8711},
    doi = {10.1111/j.1365-2966.2012.20871.x},
    url = {https://doi.org/10.1111/j.1365-2966.2012.20871.x},
    eprint = {https://academic.oup.com/mnras/article-pdf/423/1/284/18601387/mnras0423-0284.pdf},
}

@ARTICLE{Blackburne_2011,
       author = {{Blackburne}, Jeffrey A. and {Pooley}, David and {Rappaport}, Saul and {Schechter}, Paul L.},
        title = "{Sizes and Temperature Profiles of Quasar Accretion Disks from Chromatic Microlensing}",
      journal = {\apj},
         year = 2011,
        month = mar,
       volume = {729},
       number = {1},
          eid = {34},
        pages = {34},
          doi = {10.1088/0004-637X/729/1/34},
archivePrefix = {arXiv},
       eprint = {1007.1665},
 primaryClass = {astro-ph.CO},
       adsurl = {https://ui.adsabs.harvard.edu/abs/2011ApJ...729...34B}
}

@ARTICLE{Diego_2023,
       author = {{Diego}, Jose M. and {Sun}, Bangzheng and {Yan}, Haojing and {Furtak}, Lukas J. and {Zackrisson}, Erik and {Dai}, Liang and {Kelly}, Patrick and {Nonino}, Mario and {Adams}, Nathan and {Meena}, Ashish K. and {Willner}, Steven P. and {Zitrin}, Adi and {Cohen}, Seth H. and {D'Silva}, Jordan C.~J. and {Jansen}, Rolf A. and {Summers}, Jake and {Windhorst}, Rogier A. and {Coe}, Dan and {Conselice}, Christopher J. and {Driver}, Simon P. and {Frye}, Brenda and {Grogin}, Norman A. and {Koekemoer}, Anton M. and {Marshall}, Madeline A. and {Pirzkal}, Nor and {Robotham}, Aaron and {Rutkowski}, Michael J. and {Ryan}, Russell E. and {Tompkins}, Scott and {Willmer}, Christopher N.~A. and {Bhatawdekar}, Rachana},
        title = "{JWST's PEARLS: Mothra, a new kaiju star at z = 2.091 extremely magnified by MACS0416, and implications for dark matter models}",
      journal = {\aap},
         year = 2023,
        month = nov,
       volume = {679},
          eid = {A31},
        pages = {A31},
          doi = {10.1051/0004-6361/202347556},
archivePrefix = {arXiv},
       eprint = {2307.10363},
 primaryClass = {astro-ph.CO},
       adsurl = {https://ui.adsabs.harvard.edu/abs/2023A&A...679A..31D}
}

@ARTICLE{Diego_2023_gordo,
       author = {{Diego}, J.~M. and {Meena}, A.~K. and {Adams}, N.~J. and {Broadhurst}, T. and {Dai}, L. and {Coe}, D. and {Frye}, B. and {Kelly}, P. and {Koekemoer}, A.~M. and {Pascale}, M. and {Willner}, S.~P. and {Zackrisson}, E. and {Zitrin}, A. and {Windhorst}, R.~A. and {Cohen}, S.~H. and {Jansen}, R.~A. and {Summers}, J. and {Tompkins}, S. and {Conselice}, C.~J. and {Driver}, S.~P. and {Yan}, H. and {Grogin}, N. and {Marshall}, M.~A. and {Pirzkal}, N. and {Robotham}, A. and {Ryan}, R.~E. and {Willmer}, C.~N.~A. and {Bradley}, L.~D. and {Caminha}, G. and {Caputi}, K. and {Carleton}, T. and {Kamieneski}, P.},
        title = "{JWST's PEARLS: A new lens model for ACT-CL J0102{\ensuremath{-}}4915, ``El Gordo,'' and the first red supergiant star at cosmological distances discovered by JWST}",
      journal = {\aap},
         year = 2023,
        month = apr,
       volume = {672},
          eid = {A3},
        pages = {A3},
          doi = {10.1051/0004-6361/202245238},
archivePrefix = {arXiv},
       eprint = {2210.06514},
 primaryClass = {astro-ph.GA},
       adsurl = {https://ui.adsabs.harvard.edu/abs/2023A&A...672A...3D}
}

@ARTICLE{Oguri_2018,
       author = {{Oguri}, Masamune and {Diego}, Jose M. and {Kaiser}, Nick and {Kelly}, Patrick L. and {Broadhurst}, Tom},
        title = "{Understanding caustic crossings in giant arcs: Characteristic scales, event rates, and constraints on compact dark matter}",
      journal = {\prd},
         year = 2018,
        month = jan,
       volume = {97},
       number = {2},
          eid = {023518},
        pages = {023518},
          doi = {10.1103/PhysRevD.97.023518},
archivePrefix = {arXiv},
       eprint = {1710.00148},
 primaryClass = {astro-ph.CO},
       adsurl = {https://ui.adsabs.harvard.edu/abs/2018PhRvD..97b3518O}
}

@ARTICLE{Spejcher_2025,
       author = {{Spejcher}, Becca and {Richardson}, Noel D. and {Pablo}, Herbert and {Beltran}, Marina and {Butler}, Payton and {Avila}, Eddie},
        title = "{An Investigation into the Variability of Luminous Blue Variable Stars with TESS}",
      journal = {\aj},
         year = 2025,
        month = mar,
       volume = {169},
       number = {3},
          eid = {128},
        pages = {128},
          doi = {10.3847/1538-3881/ada561},
archivePrefix = {arXiv},
       eprint = {2501.00240},
 primaryClass = {astro-ph.SR},
       adsurl = {https://ui.adsabs.harvard.edu/abs/2025AJ....169..128S}
}

@ARTICLE{Ren_2019,
       author = {{Ren}, Yi and {Jiang}, Bi-Wei and {Yang}, Ming and {Gao}, Jian},
        title = "{The Period-Luminosity Relations of Red Supergiants in M33 and M31}",
      journal = {\apjs},
         year = 2019,
        month = apr,
       volume = {241},
       number = {2},
          eid = {35},
        pages = {35},
          doi = {10.3847/1538-4365/ab0825},
archivePrefix = {arXiv},
       eprint = {1902.07597},
 primaryClass = {astro-ph.SR},
       adsurl = {https://ui.adsabs.harvard.edu/abs/2019ApJS..241...35R}
}

@ARTICLE{Yang_2011,
       author = {{Yang}, Ming and {Jiang}, B.~W.},
        title = "{Red Supergiant Stars in the Large Magellanic Cloud. I. The Period-Luminosity Relation}",
      journal = {\apj},
         year = 2011,
        month = jan,
       volume = {727},
       number = {1},
          eid = {53},
        pages = {53},
          doi = {10.1088/0004-637X/727/1/53},
archivePrefix = {arXiv},
       eprint = {1011.4998},
 primaryClass = {astro-ph.SR},
       adsurl = {https://ui.adsabs.harvard.edu/abs/2011ApJ...727...53Y}
}

@misc{Sarrouh_2025,
      title={CANUCS/Technicolor Data Release 1: Imaging, Photometry, Slit Spectroscopy, and Stellar Population Parameters}, 
      author={Ghassan T. E. Sarrouh and Yoshihisa Asada and Nicholas S. Martis and Chris J. Willott and Kartheik G. Iyer and Gaël Noirot and Adam Muzzin and Marcin Sawicki and Gabriel Brammer and Guillaume Desprez and Gregor Rihtaršič and Johannes Zabl and Roberto Abraham and Maruša Bradač and René Doyon and Jacqueline Antwi-Danso and Samantha Berek and Westley Brown and Vince Estrada-Carpenter and Jeremy Favaro and Giordano Felicioni and Ben Forrest and Gaia Gaspar and Katriona M. L. Gould and Rachel Gledhill and Anishya Harshan and Nusrath Jahan and Naadiyah Jagga and Jon Judež and Danilo Marchesini and Vladan Markov and Jasleen Matharu and Shannon MacFarland and Maya Merchant and Rosa M. Mérida and Lamiya Mowla and Katherine Myers and Kiyoaki C. Omori and Camilla Pacifici and Swara Ravindranath and Luke Robbins and Victoria Strait and Visal Sok and Vivian Yun Yan Tan and Roberta Tripodi and Gillian Wilson and Sunna Withers},
      year={2025},
      eprint={2506.21685},
      archivePrefix={arXiv},
      primaryClass={astro-ph.GA},
      url={https://arxiv.org/abs/2506.21685}, 
}

@ARTICLE{2025arXiv250303829F,
       author = {{Fu}, Shuqi and {Sun}, Fengwu and {Jiang}, Linhua and {Lin}, Xiaojing and {Diego}, Jose M. and {Furtak}, Lukas J. and {Jauzac}, Mathilde and {Koekemoer}, Anton M. and {Li}, Mingyu and {Oguri}, Masamune and {Patel}, Nency R. and {Willmer}, Christopher N.~A. and {Windhorst}, Rogier A. and {Zitrin}, Adi and {Bauer}, Franz E. and {Chen}, Chian-Chou and {Chen}, Wenlei and {Cheng}, Cheng and {Conselice}, Christopher J. and {Eisenstein}, Daniel J. and {Egami}, Eiichi and {Espada}, Daniel and {Fan}, Xiaohui and {Fujimoto}, Seiji and {Hsiao}, Tiger Yu-Yang and {Jin}, Xiangyu and {Kohno}, Kotaro and {Lagattuta}, David J. and {Li}, Zihao and {Liu}, Weizhe and {Miralda-Escud{\'e}}, Jordi and {Ning}, Yuanhang and {Tacchella}, Sandro and {Tee}, Wei Leong and {Umehata}, Hideki and {Wang}, Feige and {Yan}, Haojing and {Zhu}, Yongda},
        title = "{Medium-band Astrophysics with the Grism of NIRCam In Frontier fields (MAGNIF): Spectroscopic Census of H$\alpha$ Luminosity Functions and Cosmic Star Formation at $z\sim 4.5$ and 6.3}",
      journal = {arXiv e-prints},
         year = 2025,
        month = mar,
          eid = {arXiv:2503.03829},
        pages = {arXiv:2503.03829},
          doi = {10.48550/arXiv.2503.03829},
archivePrefix = {arXiv},
       eprint = {2503.03829},
 primaryClass = {astro-ph.GA},
       adsurl = {https://ui.adsabs.harvard.edu/abs/2025arXiv250303829F}
}

@ARTICLE{Duchene_2013,
       author = {{Duch{\^e}ne}, Gaspard and {Kraus}, Adam},
        title = "{Stellar Multiplicity}",
      journal = {\araa},
         year = 2013,
        month = aug,
       volume = {51},
       number = {1},
        pages = {269-310},
          doi = {10.1146/annurev-astro-081710-102602},
archivePrefix = {arXiv},
       eprint = {1303.3028},
 primaryClass = {astro-ph.SR},
       adsurl = {https://ui.adsabs.harvard.edu/abs/2013ARA&A..51..269D}
}
\bibliographystyle{aasjournal}
\end{document}